\documentclass[preprint,aps,showpacs,groupedaddress]{revtex4}  

\usepackage{graphicx}  
\usepackage{dcolumn}   
\usepackage{bm}        
\usepackage{amssymb}   
\usepackage{booktabs}
\usepackage{textcomp}
\usepackage{color}
\usepackage{amsmath}
\usepackage{amsbsy}
\usepackage{amsfonts}
\usepackage{amstext}
\usepackage{amssymb}
\usepackage{subfigure}
\usepackage{multirow}
\usepackage{longtable}
\usepackage{lscape}
\usepackage{rotating}

\def\ii{\textrm{i}\,}

\def\ie{\textit{i.e.}}

\begin{document}

\title{Structural stability versus conformational sampling in biomolecular systems: Why is the charge transfer efficiency in G4-DNA better than in dsDNA?}

\author{P. Benjamin Woiczikowski$^{1}$}
\author{Tom\'{a}\v{s} Kuba\v{r}$^{1}$}
\author{Rafael Guti\'{e}rrez$^{2}$}
\author{Gianaurelio Cuniberti$^{2}$}
\author{Marcus Elstner$^{1}$}
\email{marcus.elstner@kit.edu}

\affiliation{$^{1}$Department for Theoretical Chemical Biology\\
Karlsruhe Institute of Technology\\
D-76131 Karlsruhe, Germany\\
}


\affiliation{$^{2}$Institute for Materials Science
and Max Bergmann Center of Biomaterials\\
Dresden University of Technology\\
D-01062 Dresden Germany\\}


\date{\today}

\begin{abstract}
The electrical conduction properties of G4-DNA are investigated using a hybrid approach, which combines electronic structure calculations, molecular dynamics (MD) simulations and the formulation of an effective tight-binding model Hamiltonian. Charge transport is studied by computing transmission functions along the MD trajectories. 
Though G4-DNA is structurally more stable than dsDNA, our results strongly suggest that the potential improvement of the electrical transport properties in the former is not necessarily related to an increased stability, but rather to the fact that G4 is able to explore in its conformational space a larger number of charge-transfer active conformations. This in turn is a result of the non-negligible inter-strand matrix elements, which allow for additional charge transport pathways. The higher structural stability of G4 can however play an important role once the molecules are contacted by electrodes. In this case, G4 may experience weaker structural distortions than dsDNA and thus preserve to a higher degree its conduction properties.
\end{abstract}

\maketitle

\section{Introduction}\label{sec:intro}

For the past two decades there has been a revival of interest in issues related to charge migration in DNA based oligomers. 
On one side, charge transfer (CT) in DNA is supposed to play a  key role in 
self-repair processes of DNA damage in natural conditions, i.e. via oxidative stress~\cite{Boon2003}. On the other side, DNA may have a huge potential application
field in the development of complex nano-scale electronic devices with self-assembling properties, since self-assembly, recognition and 
 fully-automatic synthesis of oligonucleotides
~\cite{keren03,braun98,pompe99} can allow for building almost any imaginable two and three dimensional DNA motifs
 (e.g. DNA-Origami)~\cite{Rothemund2006,Shih2004,Shih2009}. However, besides considerable progress in elucidating the possible charge transfer 
mechanisms efficient in DNA, 
there is still disagreement about the conduction properties of dsDNA, for several experiments obtained quite controversial
 results ranging from insulating~\cite{storm01}, over semi-conducting~\cite{porath00} to even metallic-like behavior
~\cite{yoo01,tao04a,cohen05}. This broad variability is related not only to the poor control of the experimental conditions in earlier transport 
experiments, but is also due to the intrinsic structural and electronic complexity of DNA oligomers (e.g. specific base pair sequence), 
which makes the formulation of a general 
theoretical framework very challenging. Nevertheless, several experiments appeared recently~\cite{Kang2010,cohen05,tao04b}
 which have shown electrical currents 
in the nA range despite the differences in the studied base sequences and in the experimental setups. One central issue which has emerged by the 
theoretical treatment of charge transport is the necessity to consider the conformational fluctuations of the molecular frame not as a weak perturbation 
but rather as a crucial factor promoting or hindering charge motion~\cite{Berlin2004,Grozema2002,Senthilkumar2005,CramerT._jp071618z,Gutierrez2009,woiczi2009}.

It has been meanwhile shown that it is possible to synthesize DNA derivatives, which do not necessarily have the double-strand structure of 
natural DNA. Thus, C and G-rich DNA strands are able to form four-stranded quadruplex structures, like the i-motif structure composed of
 two parallel hemiprotonated duplexes intercalated into each other into a head-to-tail orientation, and the G4-quadruplex 
(also known as G4-DNA) which is formed by either one, two or four G-rich DNA strands in a parallel or anti-parallel orientation~\cite{Phan2002}.
 The latter is supposed  to play an important role in some biological processes such as in telomeric DNA regions, where it 
inhibits telomerase and HIV integrase~\cite{Williamson1994rev}. Additionally, G4-DNA is known to interact with various cell proteins that cause  diseases like Bloom's and Werner's syndromes~\cite{Arthanari2001}. Moreover, G4-DNA is cytotoxic towards tumor cells and therefore, might 
be a key for the design of anti-cancer drugs~\cite{Mergny1999}. Eventually, G-quadruplexes are found to 
be thermally more stable than dsDNA~\cite{Maizels2006}. Gilbert and Sen resolved the x-ray structure of the 
first G-quadruplex and proposed the formation of these unique structures in presence of monovalent alkali ions
~\cite{Sen1988,Sen1990,Sen1992}, see also Fig.~\ref{fig:meth-tetrad}(a). Consequently, G4-DNA can be regarded as stacks composed of individual planar G-tetrads (or G-quartets) as shown in Fig.~\ref{fig:meth-tetrad}(b), each formed by 4 guanines bound together by 8 hydrogen bonds 
via Hoogsteen pairing~\cite{Hoogsteen1963}. For detailed reviews on the structural diversity and special mechanical properties 
of G4-DNA based molecules see Ref.~\cite{Davis2004,Simonsson2001}. 

Recently, Porath {\textit{et al.}}~\cite{CCohen2007} have shown the larger polarizability of G4-DNA compared to dsDNA, when attached to gold surfaces. This
 might be an indicator of improved CT properties and the potential advantage of using G4-DNA rather than dsDNA in molecular scale devices. 
Eventually, the higher conductance in G4-DNA could be attributed to the increased structural stability and higher number of overlapping
$\pi$-orbitals. Furthermore, experimental work by Kotlyar {\textit{et al.}}~\cite{Kotlyar2005,Borovok2008} revealed that up to 300 nm 
long G4-wires could be synthesized revealing a considerable stability even in the absence of metal ions. 

Also, theoretical studies have been performed on the stability and rigidity of G4-DNA. Molecular dynamics (MD)
simulations by \v{S}pa\v{c}kov\'{a} {\textit{et al.}}~\cite{Spackova1999,Spackova2001} 
could confirm the experimental results. For one thing, G4-DNA is more rigid than dsDNA and for another monovalent ions
 within the quadruplex cavity are necessary for the stability of short G4-DNA. On the
 other hand, MD simulations~\cite{Cavallari2006}  pointed out the stability of longer G4 molecules (with 24 G-tetrads)
 in absence of metal ions which is consistent with the synthetic procedure of ref~\cite{Kotlyar2005,Borovok2008}. 
Electronic structure calculations by Di Felice and co-workers~\cite{DiFelice2006book,Calzolari2004,Calzolari20023331,Calzolari2004557,DiFelice20041256,DiFelice200522301} 
showed a higher degree of delocalization of the electronic states in G4 than in dsDNA. Moreover, the presence of metal ions may contribute additional states 
supporting CT, though this is a still unresolved issue. 

Guo {\textit{et al.}}~\cite{Guo2009,Guo2010} have recently shown using the Landauer theory (coherent transport) 
that G4-DNA exhibits much larger delocalization
 lengths at the band center compared to double-stranded poly(G) DNA. Surprisingly, it was also found out that the
 delocalization length can be even enhanced via environment-induced disorder through the backbones. 
Though these studies assume a static atomic structure, they nevertheless
suggest that disorder plays an important, non-trivial role in mediating CT in DNA.

Similarly, it is known from other biomolecules such as proteins that electron transport is dominated by 
non-equilibrium fluctuations which has lately been analyzed by Balabin {\textit{et al.}}~\cite{Skourtis2008}. 
As demonstrated in previous studies, idealized static structures are not representative when considering CT properties,  
since dynamical as well as environmental effects were shown to be too important
~\cite{Voityuk2004ACIE,Berlin2008JPCC,Voityuk2006CPL,Voityuk2007CPL,Grozema2008}. It was indicated that 
dynamic disorder has dramatic effects, since it suppresses CT in homogeneous sequences on the one hand,
 but can  enhance CT in heterogeneous  sequences, on the other. For instance, in our previous work~\cite{woiczi2009},
 the conductance of the Dickerson dodecamer (sequence: 5'-CGCGAATTCGCG-3') was found to be almost one order of 
magnitude larger in solution (QM/MM) than in vacuo. Interestingly, similar findings have recently
 been shown by Scheer and co-workers, for they obtained for a heterogeneous sequence (i.e. A and G bases are present) with 31 base pairs
~\cite{Scheer-Sequence} a two order of magnitude larger conductance 
in solution compared to the in vacuo measurements~\cite{Kang2010}. Furthermore, it could be shown that only a 
minor part of the conformations is CT-active~\cite{woiczi2009}. Therefore, neglecting these significant factors or assuming a purely random disorder distribution can lead to a considerable loss of a vital part of CT-relevant structural and electronic information.

Relying on our previously developed methodologies to deal with CT in different DNA oligomers~\cite{woiczi2009,Gutierrez2009,Gutierrez2010}, we present here a detailed investigation of charge transport in G4-DNA.

The paper is organized as follows, in Sec.~\ref{sec:meth} details of the MD simulations and of the electronic structure approach are described. Subsequently, the MD and  the electronic structure data are analyzed in Sec.~\ref{subsec:res-md} and \ref{subsec:res-es}, respectively. Finally in Sec.~\ref{subsec:res-landauer}, CT results for G4 molecules are compared with those of dsDNA in solution and in vacuo. More importantly, we point out significant factors which are responsible for the enhanced conductance in G4 molecular wires.

\section{Methodology}\label{sec:meth}

\subsection{Starting Structures and Simulation Set-Up}\label{subsec:meth-md}

\begin{figure}
\includegraphics{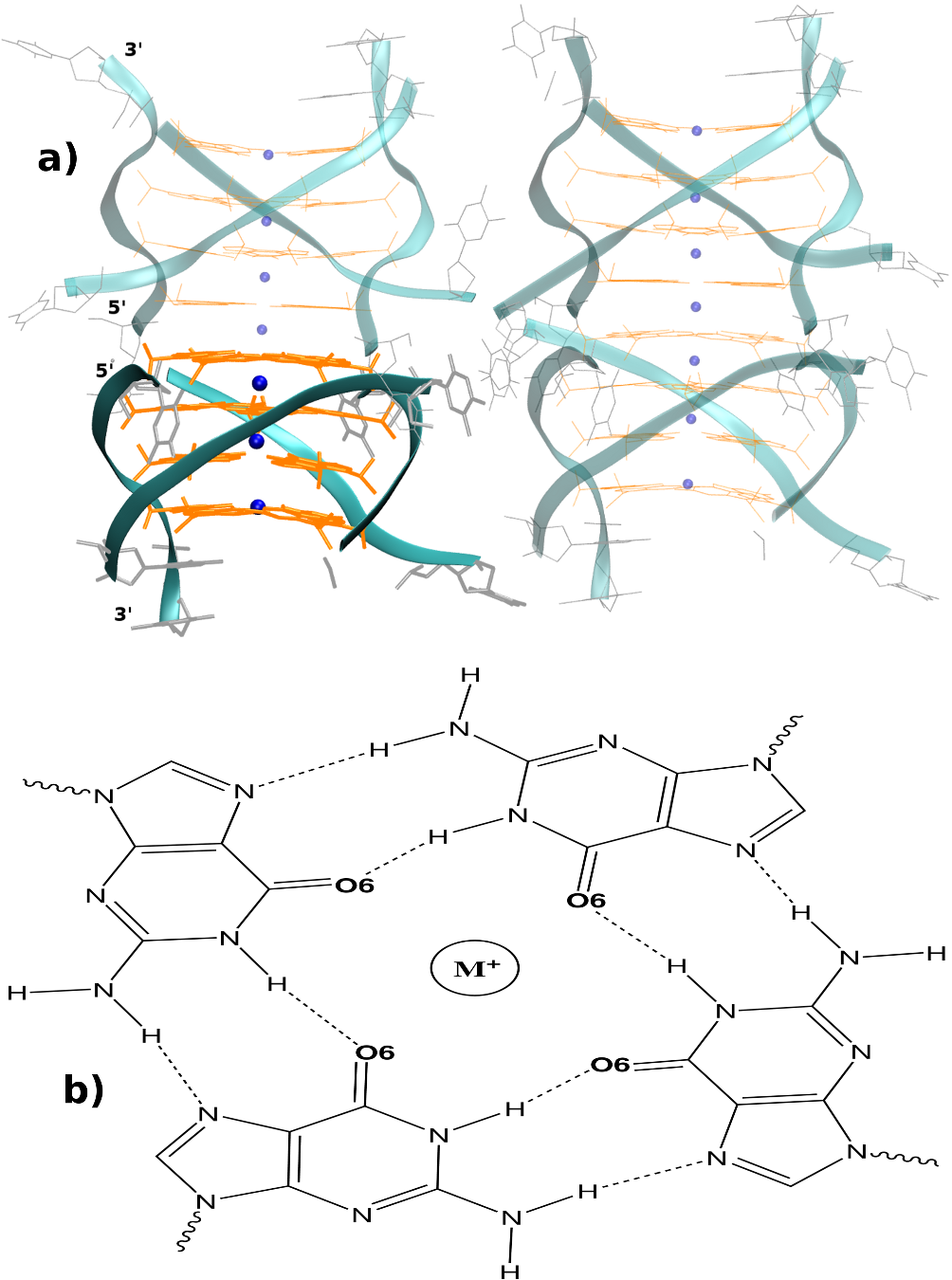}
\caption{a) Tetrameric G4 x-ray crystal structure (244d~\cite{Laughlan1994}) with four units (TG$_4$T)$_4$. Backbones are indicated as light blue ribbons, central coordinated sodium ions as dark blue spheres, terminal thymine and guanine residues are shown in gray and orange, respectively. The highlighted quadruplex is used as starting structure and also to generate longer G4 molecules (G$_{12}$)$_4$ and (G$_{30}$)$_4$. b) Lewis structure of a single G4-tetrad with C$_{4h}$ symmetry containing a monovalent metal ion in its center. Guanines are bound together by 8 hydrogen bonds via Hoogsteen pairing~\cite{Hoogsteen1963}. The metal ion is coordinated either coplanar by 4 or cubic by 8 O6 oxygen atoms of the respective guanines which depends on the ionic radius of M$^+$.}\label{fig:meth-tetrad}%
\end{figure}

The molecules used in this work are based on the x-ray crystal structure of a tetrameric 
parallel-stranded quadruplex (244d) formed by the hexanucleotide sequence d(TG$_4$T) in the presence of sodium ions. 
The structure has been resolved at 1.2 \AA{} resolution by Laughlan {\textit{et al.}}~\cite{Laughlan1994}. 
As shown in Fig.~\ref{fig:meth-tetrad}(a), it contains 2 pairs of quadruplexes (TG$_4$T)$_4$, for which each
 pair is stacked coaxially with opposite polarity at the 5' ends. Moreover, 9 sodium ions are located inside 
each stack of quadruplexes illustrating  well ordered G4-DNA constructs. However, the terminal thymine residues, 
shown in gray, were not completely resolved because of high thermal disorder~\cite{Laughlan1994}. As indicated 
in Fig.~\ref{fig:meth-tetrad} only one of these 4 quadruplexes is taken as our basis structure, which corresponds 
to the parallel strands A, B, C and D from the PDB file 244d. This quadruplex, from now on denoted as (TG$_4$T)$_4$, 
will be used for simulations and calculations. Furthermore, two longer G4 quadruplexes with 12 and 30 tetrads, denoted 
as (G$_{12}$)$_4$ and (G$_{30}$)$_4$, are generated by omitting the terminal thymine residues and adding subsequently 
G4 tetrads with a distance of 3.4 \AA{} and twisted by 30\textdegree{}. For comparison, corresponding double-stranded 
B-DNAs with base sequence 5'-TGGGGT-3', poly(G) denoted as G$_{12}$ and G$_{30}$, respectively, and a sequence 
containing 31 base pairs~\cite{Scheer-Sequence} (Scheer) are built with the make-na server~\cite{link00}.

Depending on the helix length the G4 and dsDNA molecules are solvated in a rectangular box with
  4000 to 11000 water molecules using the TIP3P model~\cite{TIP3P} and periodic boundary conditions
 (PBC) are applied. In order to neutralize the total charge of the system, due to the negatively charged backbones, 
an appropriate number of sodium counterions (Na+) are added. Concerning the quadruplexes, we carried out 4 simulations 
for (TG$_4$T)$_4$ and (G$_{12}$)$_{4}$, respectively, firstly without any central monovalent alkali ions within the quadruplex, 
and then  in presence of either lithium (Li+),  sodium (Na+), or potassium (K+) ions within the respective G4 molecules. 
Therefore, either 3 ((TG$_4$T)$_4$) or 11 ((G$_{12}$)$_{4}$) of these ions are subsequently placed in the center between two G4 tetrads coordinated by O6 oxygen 
atoms (see also Fig.~\ref{fig:meth-tetrad}(b)). Nevertheless, the longer (G$_{30}$)$_4$ quadruplex is simulated  only twice in
absence and presence of central sodium ions. All simulations are carried out with the GROMACS software package~\cite{Gromacs} using the 
AMBER parm99 forcefield~\cite{parm99} 
including the parmBSC0 extension~\cite{parmBSC0}. After a standard heating-minimization protocol followed by a 1 ns 
equilibration phase, which is discarded afterwards, we performed 30 ns (50 ns for (G$_{30}$)$_4$) MD simulations with 
a time step of 2 fs. Snapshots of the molecular structures were saved every 1 ps, for which the charge transfer 
parameters were calculated with the SCC-DFTB-FO approach as described in Sec.~\ref{subsec:meth-hop}.  

\subsection{Electronic structure}\label{subsec:meth-hop}

In this section we will shortly describe how the electronic structure of G4-DNA molecules is mapped to 
a coarse-grained transfer Hamiltonian using the fragment orbital (FO) approach. The method has already 
successfully been applied to dsDNA molecules~\cite{Kubar2008b,woiczi2009,Gutierrez2009,Gutierrez2010}. A more detailed description of
 the methodology can be found in Ref.~\cite{Kubar2008a}. To begin with, the electronic structure of the DNA system is written in 
an effective tight-binding  basis (fragment basis) as:   
\begin{equation}
H = \sum_i \varepsilon_i a_i^{\dagger} a_i + \sum_{ij} T_{ij} (a_i^{\dagger}  a_j + \textrm{h.c.}).
\end{equation}
The onsite energies $\varepsilon_i$ and the nearest-neighbor hopping integrals $T_{ij}$ characterize,
 respectively, effective ionization energies and electronic couplings of the molecular fragments . 
The evaluation of these parameters can be done very efficiently using the SCC-DFTB method~\cite{ElstnerPRB1998} 
combined with a fragment orbital (FO) approach~\cite{Senthilkumar2005}:
\begin{equation}
\label{eq:eps} \varepsilon_i=-\left<  \phi_i   \left|   \hat{H}_{\mathrm{KS}}          \right| \phi_i
\right>
\end{equation}
 and 
\begin{equation}
\label{eq:Tij}
T^{0}_{ij}=\left<  \phi_i   \left|   \hat{H}_{\mathrm{KS}}  \right| \phi_j \right>.
\end{equation}
The indexes $i$ and $j$ correspond to the fragments of the quadruplexes which, in this model, are 
constituted by single guanine bases. Accordingly, the molecular orbitals $\phi_i$ and $\phi_j$ 
 are the respective highest-occupied molecular orbitals which are obtained by performing SCC-DFTB calculations
 for these isolated fragments, i.e. the individual guanine bases. Using a linear combinations of atomic orbitals ansatz $\phi_i=\sum\limits_\mu{c_\mu^i\eta_\mu}$
 the coupling and overlap integrals can be efficiently evaluated as: 
\begin{equation}\label{eq:Tij lcao}
 T^{0}_{ij}=\sum\limits_{\mu\nu}c_\mu^ic_\nu^j\left< \eta_\mu \left| \hat{H}_{\mathrm{KS}} \right| \eta_\nu \right>
 = \sum\limits_{\mu\nu}c_\mu^ic_\nu^j H_{\mu \nu}
\end{equation}
and
\begin{equation}\label{eq:Sij_lcao}
 S_{ij}=\sum\limits_{\mu\nu}c_\mu^ic_\nu^j\left<  \eta_\mu   \middle|  \eta_\nu \right> =
 \sum\limits_{\mu\nu}c_\mu^ic_\nu^j S_{\mu \nu}.
\end{equation}
$H_{\mu \nu}$ and $S_{\mu \nu}$ are the Hamilton and overlap matrices in the atomic
basis set as evaluated with the SCC-DFTB method. Note, since $T^{0}_{ij}$ is built from non-orthogonal orbitals
 $\phi_i$ and $\phi_j$, which is for various problems not suitable, we apply the Loewdin transformation~\cite{LoewdinJCP1950}:
\begin{equation}
\mathbf{ T=S^{-\frac{1}{2}}T^{0}S^{-\frac{1}{2}}}.
\end{equation} 
The effect of the environment, \ie~the electrostatic field of the DNA backbone, the water molecules and the
 counter-ions (including the central alkali ions within the quadruplex), is taken into account through 
the following QM/MM Hamiltonian:
\begin{equation}
\begin{split}
 H_{\mu\nu} = H_{\mu\nu}^0 & +\frac{1}{2} S_{\mu \nu}^{\alpha
\beta} \Bigg( \sum_{\delta} \Delta q_{\delta}(\gamma_{\alpha \delta}+\gamma_{\beta \delta})\\
& + \sum\limits_A Q_A \left(\frac{1}{r_{A\alpha}}+ \frac{1}{r_{A\beta}}\right) \Bigg)\\
\end{split}
\label{eq:QMMM}
\end{equation}
$\Delta q_{\delta}$ are the Mulliken charges in the QM region and $Q_A$ are the charges in the MM region, 
\ie~the DNA backbone, counterions and water molecules. The coupling to the environment is therefore explicitly 
described via the interactions with the $Q_A$ charges. In the following, we will denote the calculation set-up 
based on the complete expression in Eq.~\ref{eq:QMMM} as QM/MM;  neglecting the last term will be denoted as ``vacuo''.

The electronic parameters $\varepsilon_i$ and $T_{ij}$ are evaluated for every snapshot of the simulations. 
Subsequently, the transfer Hamiltonian is used to calculate the CT in the quadruplexes as described in Sec.~\ref{subsec:meth-tc}.

\subsection{Charge transport through a 4-stranded quadruplex}\label{subsec:meth-tc}
Once the transfer Hamiltonian has been constructed, the transport properties will be calculated using Landauer theory for 
each snapshot along the MD trajectories. For this, we consider a two-terminal set up where the G4 oligomer is contacted to left (L) and right (R) 
metallic electrodes through the four terminal bases. A central quantity to be computed is the  molecular (G4) Green function, which can be 
obtained through a Dyson equation:
\begin{eqnarray}
 \mathbf{{G}^{-1}}(E)&=&E\bf{1}-\bf{H}-\bf{\Sigma}_{\textrm{L}}-\bf{\Sigma}_{\textrm{R}}.
\end{eqnarray}
Here $\Sigma_L$ and $\Sigma_R$ are self-energy matrices characterizing the coupling to the electrodes. To simplify the calculations
we do not take the full energy dependence of the self-energies into account, but rather use the so called wide band limit, where $\Sigma_L$ and $\Sigma_R$ 
are replaced by energy-independent parameters~\cite{mujica:6849,mujica:6856}: 
\begin{eqnarray}
  ({\bf{\Sigma}}_{\textrm{L}})_{lj}&=&-\ii\gamma_{\textrm{L}}\delta_{lk}\delta_{jk}, (k=1,2,3,4)\nonumber \\
 ({\bf{\Sigma}}_{\textrm{R}})_{lj}&=&-\ii\gamma_{\textrm{R}}\delta_{lk}\delta_{jk}, (k=N-3,N-2,N-1,N). \nonumber
\end{eqnarray}
We set $\gamma_{\textrm{L}}$ and $\gamma_{\textrm{R}}$ to 1 meV. Within the Landauer approach, 
the transmission function $T(E)$ for a given set of electronic parameters is then obtained as:
\begin{eqnarray}
 T(E) = Tr\left[{\bf{\Gamma}}_{\textrm{L}} {\bf G} {\bf{\Gamma}}_{\textrm{L}} {\bf G^+} \right]
\end{eqnarray}
where ${\bf{\Gamma}}_{\textrm{L}}$ and ${\bf{\Gamma}}_{\textrm{L}}$ are the broadening matrices calculated as the anti-hermitian part of $\bf{\Sigma}_{\textrm{L/R}}$:
\begin{eqnarray}
{\bf{\Gamma}}_{\textrm{L/R}} = \ii \left[\bf{\Sigma}_{\textrm{L/R}}-\bf{\Sigma}_{\textrm{L/R}}^+ \right]
\end{eqnarray}  
Using the former expressions, the conformational (time) dependent electrical current will be simply calculated by: 
\begin{eqnarray}
 I(U,t)&=&\frac{2e}{h}\int dE\,\Bigg( f \bigg(E-E_{\textrm{F}}-\frac{eU}{2}\bigg)\nonumber \\&-&f\bigg(E-E_{\textrm{F}}+\frac{eU}{2}\bigg)\Bigg) \, T(E,t)
\end{eqnarray}
The I-U characteristics presented in this work should be interpreted only qualitatively, 
for the Fermi energy ($E_\textrm{F}$) is artificially placed as average of the onsite energies for 
all guanine sites and for each snapshot, respectively. The reader should note that the current could 
exhibit quite different shapes depending on where $E_\textrm{F}$ is located. However, only the current-voltage gap is affected by $E_\textrm{F}$, whereas the maximum current obtained 
at high voltages is not altered.

The transmission function and the current are evaluated for every snapshot of the MD simulation in order to obtain statistical average quantities. However, we are aware of the limits of the coherent transport model which is valid only in the adiabatic regime. Hence, we assume that the timescales of the CT process is shorter  than the fastest dominant structural fluctuations. This issue has also been addressed  in Ref.~\cite{woiczi2009}.

\section{Results}

\subsection{Structural stability and rigidity in parallel stranded quadruplexes}
\label{subsec:res-md}

\begin{figure}\centering
 \includegraphics{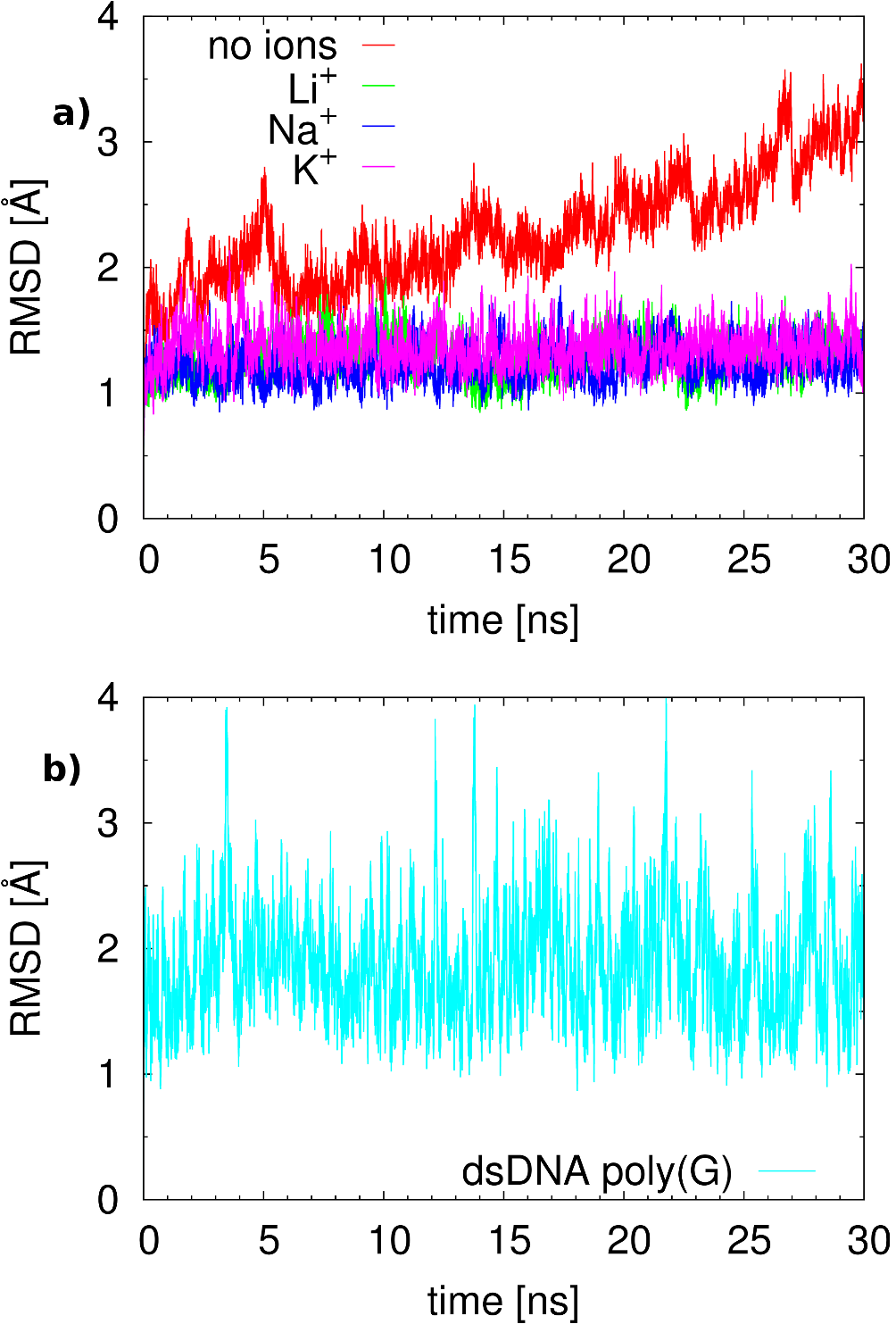}
\caption{RMSD of G4 and poly(G) DNA: a) parallel stranded quadruplex (G$_{12}$)$_4$ in absence and presence of centrals ions Li$^+$, Na$^+$ and K$^+$, b) comparison with double-stranded G$_{12}$.}\label{fig:rmsd2}
\end{figure}

\begin{figure}\centering
 \includegraphics{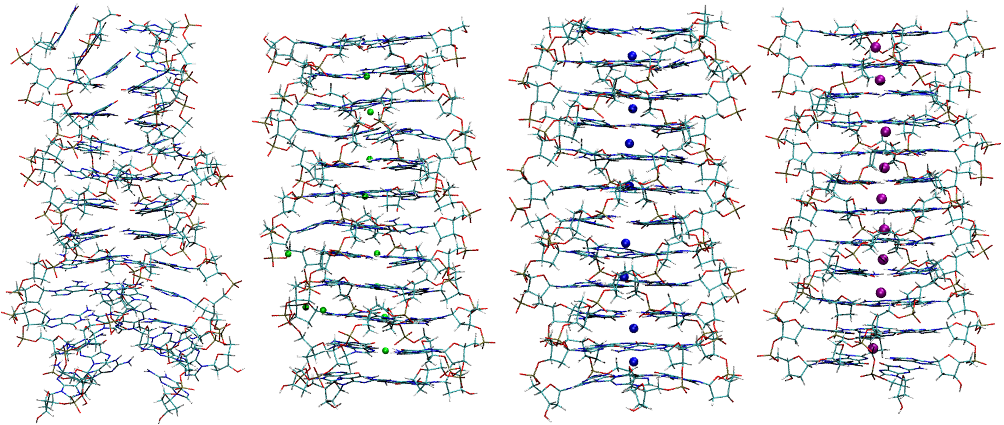}
\caption{Molecular snapshot of parallel stranded quadruplex (G$_{12}$)$_4$ after 30 ns MD simulation in absence and presence of centrals ions Li$^+$ (green spheres), Na$^+$ (blue spheres) and K$^+$ (purple spheres).}\label{fig:strucs-12g4}
\end{figure}

First of all, the (TG$_4$T)$_4$ quadruplex from the crystal structure 244d has been simulated in order 
to validate further simulations with 12 and 30 tetrads. Corresponding RMSD time series in absence and presence of central ions as well as molecular snapshots after 30 ns simulation time can be found in Sec.~SI.A. in Ref.~\cite{supp}. Based on the crystal structure 244d we built up an artificial all-parallel stranded guanine quadruplex (G$_{12}$)$_{4}$ without terminal thymine residues and 12 stacked guanine tetrads (for details see~\ref{subsec:meth-md}). As for (TG$_4$T)$_4$, we carried out 4 simulations, one in absence and three in presence of 11 lithium, sodium and potassium ions initially placed in the center of two adjacent G4 tetrads. The RMSD time series and molecular snapshots after 30 ns MD simulation for (G$_{12}$)$_{4}$ and poly(G) (G$_{12}$) are given in Fig.~\ref{fig:rmsd2} and Fig.~\ref{fig:strucs-12g4}, respectively.

The RMSD values show that without central ions the quadruplex cannot be considered as a G4 structure in equilibrium. In fact, the RMSD increases for the whole simulation time from 1.5 to 3 \AA{} indicating considerable structural changes. In contrast, the RMSD for structures containing central ions reveal rather consistent values around 1 to 1.4
 \AA{} with only minor fluctuations suggesting rigid G4 structures with only slight structural deviations. Compared to
 the (TG$_4$T)$_4$ simulations, the type of ions within the quadruplex has a lesser influence on the structural 
deviations, since the RMSD values are rather similar. In order to validate these small structural deviations and therefore 
the rigidity of G4-DNA we carried out a RMSD calculation for double-stranded G$_{12}$ (Fig.~\ref{fig:rmsd2}(b)). 
Generally, the RMSD values and their fluctuations for the dsDNA molecule are substantially larger than those for
 the quadruplex structures. This clearly supports the notion that guanine quadruplexes are much more rigid than 
corresponding dsDNA and therefore G4-DNA reveals less dynamical disorder which might promote CT.

The snapshots in Fig.~\ref{fig:strucs-12g4} exhibit that (G$_{12}$)$_{4}$ in absence of central ions is not completely destabilized which has not been the case for (TG$_4$T)$_4$ (see Sec.~SI.A. in Ref.~\cite{supp}). The inner tetrads roughly maintain a regular G4 structure, whereas the outer ones show a considerable disordered behavior with no 
planarity at all, even the G4 hydrogen bridge pattern has been entirely destroyed. Recently, Cavallari {\textit{et al.}} have
 shown in their MD simulations that long G4 quadruplexes can be stable without central ions~\cite{Cavallari2006}. 
This finding was supported by experimental results in which longer G4 quadruplexes could be synthesized and characterized 
in absence of monovalent ions~\cite{Kotlyar2005}. 

Therefore, an additional 50 ns simulation of a quadruplex with 30 tetrads (G$_{30}$)$_4$ in absence of central ions 
is carried out. The simulations clearly exhibit that the major part of the quadruplex forms a quite regular structure
 with adequate parallel stacking. Therefore, it seems to confirm previous results that large G4 wires in absence of
 central ions become more stable, reasonably due to increased stacking interaction. However, like for the dodecamer,
 towards the ends the quadruplex is slightly deformed, especially at the 3' terminus. Moreover, RMSD values reveal that 
the quadruplex is not completely structural balanced. The major part of structural fluctuation originates from the 5' and 3'
 ends which is shown by four groups of large peaks in root mean square fluctuations per atom (RMSF). The corresponding data is shown
 in Fig.~S2 in Ref.~\cite{supp}.
 
The quadruplexes which contain central monovalent ions in Fig.~\ref{fig:strucs-12g4} exhibit highly regular G4 structures indicating again the strong rigidity and stability 
of these molecules. Generally, the ions appear to prefer different coordinations within the quadruplex cavity, that is lithium favors the planar position within the tetrads while sodium and potassium are most likely to be found in the center of two G4 planes. However, one sodium ion in the middle of the quadruplex is located in-plane and two others are rather closer  to one of the corresponding tetrads than sandwiched between them. Also, lithium ions are sometimes found slightly apart from the in-plane position. This might be an indicator for the longitudinal mobility of central sodium and lithium ions within guanine quadruplexes. In contrast, potassium ions are rather fixed in their center position between two tetrads, most likely  because of their larger ionic radii compared with lithium and sodium. The resulting larger rigidity of G4 compared to dsDNA plus the preferred locations of central ions within the quadruplex are in perfect agreement with findings obtained by \v{S}pa\v{c}kov\'{a} {\textit{et al.}} as well as previous work by Cavallari {\textit{et al.}}~\cite{Spackova1999,Spackova2004,Cavallari2006}.

Notwithstanding, the ions are equally capable of moving out of the quadruplex through the minor groove and the 5' and 3' ends. 
In our simulations 3 sodium, 3 lithium and 2 potassium ions went out of the G4 dodecamer, respectively. Sometimes, these ions 
remain in the minor groove for several ps. Likewise, solvent molecules are able to infiltrate the quadruplex via the minor 
groove and the 5' and 3' ends provided that these locations are not already occupied by initially placed ions. Usually, the 
infiltration of solvent molecules into the helix will not significantly destabilize the structure as long as there is at least
 a minimum amount of central monovalent cations left in the quadruplex. Otherwise, the interior solvation might destroy the regular
 4-stranded structure as can be seen for the structures without central ions in Fig.~\ref{fig:strucs-12g4}. These observations are consistent with results of previous 
works~\cite{Cavallari2006,Spackova1999,Spackova2004}.

\subsection{Electronic Structure of G4 Tetrads}\label{subsec:res-es}
\subsubsection{Idealized Model}
Before calculating CT characteristics, we have to focus on the electronic parameters, 
especially the molecular orbitals involved in the hole transfer process. Similarly 
to our previous work on CT parameters in dsDNA~\cite{Kubar2008a} we carried out benchmark 
calculation of G4 tetrads with DFTB and compared to DFT (PBE and B3LYP) as well as HF calculation. 
Since we intend to deal with hole transfer we are most notably interested in the highest molecular 
orbitals which ought to have $\pi$-symmetry in order to have sufficient MO overlap along the quadruplex. 
Tab.~\ref{tab:MO-ideal} shows energies of the 5 highest occupied MOs for a idealized G4 structure with 
C$_{4h}$-symmetry obtained after geometry optimization with B3LYP/6-31G(d,p). Corresponding visualization 
of MOs can be found in Tab.~S3 in Ref.~\cite{supp}. 

\begin{table}\centering
\caption{Energies of highest occupied molecular orbitals for a idealized G4 tetrad with C$_{4h}$-symmetry, 
comparison between DFTB, DFT and HF, for the latter the 6-31G(d,p) basis set is used, all values in eV.}\label{tab:MO-ideal}
\begin{tabular}{lccccc}
\toprule[1pt]
Method   &	HOMO   & HOMO-1 & HOMO-2 & HOMO-3 & HOMO-4 \\
\midrule[.5pt] 
DFTB	 &	-4.588 & -4.593 & -4.593 & -4.597 & -4.870  \\
PBE	 &	-4.356 & -4.383 & -4.383 & -4.408 & -5.267  \\
B3LYP	 &	-5.170 & -5.196 & -5.196 & -5.222 & -6.637  \\
HF 	 &	-7.777 & -7.803 & -7.803 & -7.828 & -10.827 \\
\bottomrule[1.pt]
\end{tabular}
\end{table}

The first four occupied 
MOs, HOMO to HOMO-3 are very close in energy for all methods respectively and they have $\pi$-symmetry. The range is about 0.01 eV for DFTB and 0.05 eV for the rest. In contrast the HOMO-4 is separated significantly by a gap of about 0.3 eV for DFTB and at least more than 0.8 eV for the other methods. For DFT based methods this MO has $\sigma$-symmetry. As a result HOMO to HOMO-3 are the orbitals used for the transport calculations, for the close energy range between them supports the notion of band-like transport provided large G4 stacks are considered. By contrast, static disorder in dsDNA due to differences in ionization potentials between 
four different nucleotides (A,C,G and T) leads to large energy gaps for tunneling and therefore to rather 
localized states. Basically DFTB MO energies lie between those for PBE and B3LYP.

More importantly, these MOs appear to be linear combinations of HOMOs for the isolated guanine molecules which can be shown by comparison with our previous results~\cite{Kubar2008a}. Therefore, we are able to map the electronic structure onto single guanines rather than onto whole tetrads, since this reduces the computational costs immensely.

In Sec.~II.B. in Ref.~\cite{supp}, we discuss the effect of central ions on the MOs of G4-tetrads both purely electrostatically and quantum mechanically. As a main result, we find that the MO's are shifted down due to the electrostatic interactions with the ions. This effect is captured by a classical treatment of the ions, therefore, in our MD simulations we can treat the ions with MM rather than with QM.

\subsubsection{Dynamics and QM/MM}

Up to now we only calculated ideal G4 tetrads which where highly symmetric, therefore we applied the previous 
routine on a molecular snapshot taken from a MD simulation. That way we can analyze the effects of dynamics 
as well as the purely electrostatic environment built of backbone and solvent charges on the electronic structure 
of G4 tetrads. As shown in Sec.~II.B in Ref.~\cite{supp} single alkali ions placed in the quadruplex cavity possess a substantial impact
 on the MO energies of hole transfer states. Therefore, it is vital to include the MM environment in our calculations. 
The MO energies in Tab.~\ref{tab:MO-qmmm} for all applied methods are shifted down by about 0.5 to 0.8 eV compared to those 
for the idealized tetrad without MM environment. Also, the energy range in which HOMO to HOMO-3 are located has become 
considerably larger from 0.01 to about 0.15 eV for DFTB and equally from 0.05 to almost 0.4 eV for DFT and HF. Accordingly, 
these hole transfer states become rather localized onto the single guanines than onto the whole tetrads as soon as the
 electrostatic environment is considered, although the energy range is still very small. 

\begin{table}\centering
\caption{Energies of highest occupied molecular orbitals of a single tetrad of (TG$_4$T)$_4$ for a MD snapshot containing the
 full electrostatic environment built of negatively charged backbone, solvent and counterions. Comparison between DFTB, DFT and 
HF, for the latter the 6-31G(d,p) basis set is used, all values in eV.}\label{tab:MO-qmmm}
\begin{tabular}{lccccc}
\toprule[1pt]
Method & HOMO & HOMO-1 & HOMO-2 & HOMO-3 & HOMO-4 \\  
\midrule[.5pt]
 DFTB 	& -5.390 & -5.409 & -5.467 & -5.544 & -5.756 \\
 PBE  	& -5.052 & -5.156 & -5.206 & -5.273 & -5.986 \\
 B3LYP	& -5.871 & -5.949 & -5.994 & -6.121 & -7.356 \\
 HF	& -8.322 & -8.328 & -8.362 & -8.711 & -11.113\\
\bottomrule[1.pt]
\end{tabular}
\end{table}  

In Tab.~S5 in Ref.~\cite{supp} we provide snapshots of corresponding MOs for these states. However, this effect of localization clearly supports the notion to build up a transport model with single guanines as fragments rather than whole tetrads. It has not only lower computational costs, for only 16  instead of 64 atoms have to be calculated per fragment, but it also offers a large flexibility because CT can occur through a multitude of pathways along the quadruplex.

\subsection{Electronic Parameters}

\begin{figure}%
\includegraphics{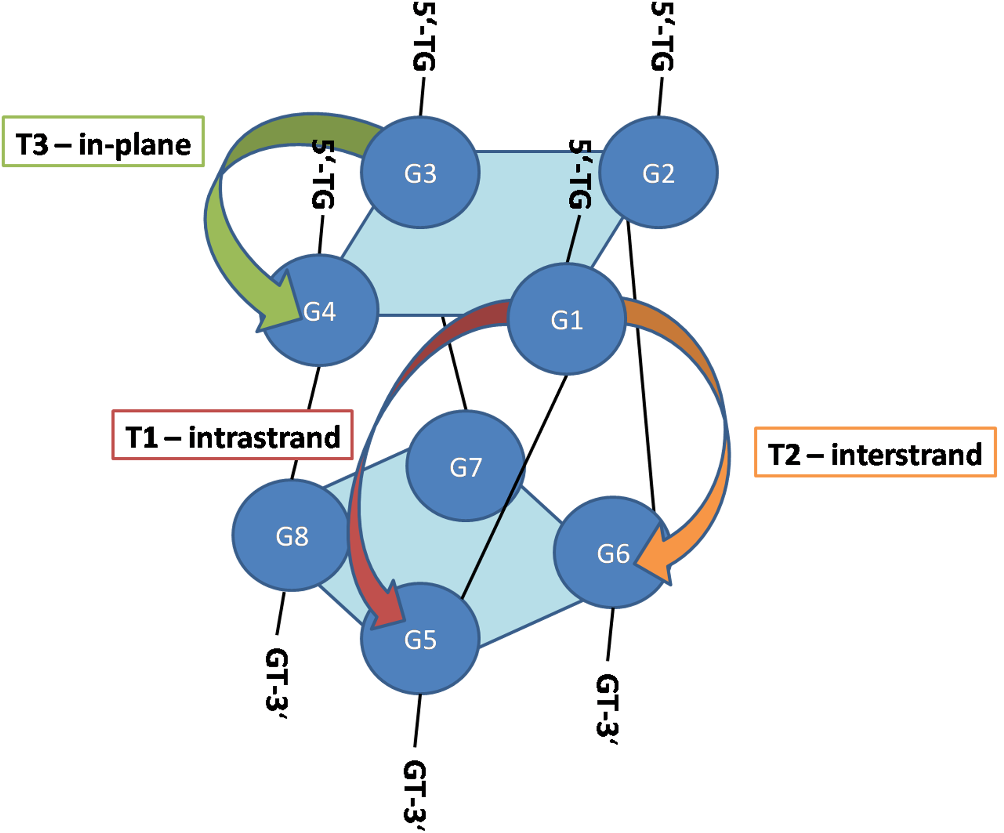}%
\caption{Scheme for electronic couplings T$_{ij}$ in G4-DNA shown for the two inner most tetrads in (TG$_4$T)$_4$. Representatively, T1 indicates intrastrand, T2 interstrand and T3 in-plane couplings.}%
\label{fig:hop-scheme}%
\end{figure}

In this section we analyze CT parameters obtained for the simulations of (TG$_4$T)$_4$ and compare them to those for the crystal structure 244D as well as for an idealized G4 stack. The parameters were calculated as described in Sec.~\ref{subsec:meth-hop}. Before showing the results, first the applied fragment methodology is introduced. The scheme in Fig.~\ref{fig:hop-scheme} shows different types of couplings present in G4-DNA. Here T3 represents electronic couplings within the respective G4-tetrads in-plane. Whereas T1 and T2 denote intra- and interstrand couplings lengthways to the quadruplexes which occur on either one strand or between two strands, respectively. In contrast to dsDNA, experiments suggest that also competing horizontal CT can occur in G4-DNA~\cite{Ndlebe2006}.

Clearly, a charge can follow several pathways along the quadruplex strongly dependent on those three couplings. This should be an advantage compared to dsDNA. Even in small G4 stacks there is a large number of electronic couplings. We will see that only a minor part of them will be vital for CT in G4-DNA. For instance, T3 couplings are usually quite small, especially the diagonal in-plane couplings in Fig.~\ref{fig:hop-scheme} (G1-G3,G2-G4, ...) are generally negligible. Also, 
most of T2 couplings are small, except those between adjacent strands (G1-G6,G2-G7, ...), since these are rather close to each other. The largest couplings are certainly found for T1 (intrastrand) and they are well comparable  to those for double-stranded poly(G) DNA~\cite{Kubar2008a}.

A summary of the average electronic couplings and onsite energies for the simulations of (TG$_4$T)$_4$ compared to the crystal structure (244D) as well as to the corresponding dsDNA, i.e. the two central guanines in 5'-TGGGGT-3' is given in Tab.~\ref{tab:hop-g4}.

\begin{table*}\centering
\caption{Average onsite energies $\left<\varepsilon\right>$ and electronic couplings T1, T2 and T3 with standard deviations for the two central guanine tetrads of (TG$_4$T)$_4$. Energies obtained from MD simulations in absence and presence of ions are compared to the crystal structure (244D), an idealized G4 dimer stack, and also to the corresponding dsDNA structure, i.e. the two central guanines in 5'-TGGGGT-3'. Note that averaging is carried out not only along the MD time series but also over 8 and 2 G bases for G4 and dsDNA, respectively. All values in eV.}\label{tab:hop-g4}
\begin{tabular}{lccccccc}
\toprule[1pt]
Type & ideal & 244D & MD no ions & MD Li+ & MD Na+ & MD K+ & dsDNA \\
\midrule[.5pt]
$\varepsilon$ & -4.895 & -4.905$\pm$0.062 & -4.812$\pm$0.368 & -5.339$\pm$0.370 & -5.400$\pm$0.350 & -5.201$\pm$0.354 & -4.790$\pm$0.371\\
T1	   &  0.028 &  0.051$\pm$0.011 &  0.039$\pm$0.028 &  0.031$\pm$0.021 &  0.031$\pm$0.021 &  0.029$\pm$0.020 & 0.052$\pm$0.034 \\
T2 	   &  0.001 &  0.012$\pm$0.002 &  0.010$\pm$0.013 &  0.022$\pm$0.014 &  0.015$\pm$0.012 &  0.013$\pm$0.010 & 0.004$\pm$0.005\footnote{interstrand coupling for G\textbackslash C, G/C=0.013$\pm$0.014 eV} \\
T3 	   &  0.009 &  0.007$\pm$0.003 &  0.009$\pm$0.009 &  0.007$\pm$0.005 &  0.006$\pm$0.004 &  0.007$\pm$0.004 & 0.012$\pm$0.008\footnote{Coupling within the WCP between G and C, note there is an energy gap of 0.4 eV} \\
\bottomrule[1.pt]
\end{tabular}
\end{table*} 

As has been discussed above, the central ions stabilize the G4 structure significantly, the structural fluctuations of G4 with ions are much lower compared to G4 without ions or dsDNA. The effect of ions on the electronic parameters is twofold: First of all,
 the onsite energies
are shifted down by about 0.5 eV but , surprisingly, the enhanced structural stability does not lead to smaller fluctuations of 
the onsite parameters. They
are in the order of 0.4 eV, similar to the situations in dsDNA, as discussed recently~\cite{Kubar2008b}. These large parameter
 variations are introduced
by solvent fluctuations (and not by fluctuations of the DNA/G4 structure itself), therefore, they are not affected by the higher
 structural rigidity.
The T1 values, on the other hand, are even lower in the G4 structures with ions. This would indicate, that these structures conduct
even less, when compared to four strands of dsDNA. However, the T2 and T3 values in G4 are still large, indicating that 
interstrand transfer can occur quite frequently. This opens a multitude of pathways for charge transport in G4 (compared to dsDNA),
which will be the key to understand G4 conductivity, as discussed in more detail below.

As a first result, we do not see any indication that the higher structural stability of G4 with central ions will lead to a higher conductivity due to a reduced dynamical onsite disorder or due to increased electronic couplings, i.e. because of somehow better stacking interactions owing to the more regular structure. Therefore, the higher conductivity must have different reasons. The more stable structure leads to smaller couplings, in contrast to prior expectations \cite{CCohen2007}. A more detailed analysis of onsite energies and electronic couplings is given in Tab.~S6-S11 in Ref.~\cite{supp}.

\subsection{Coherent Transport in G4-DNA}\label{subsec:res-landauer}

The parameters $\varepsilon$ and T$_{ij}$ analyzed in the previous section are now used to evaluate the charge transport properties using the approach described in Sec.~\ref{subsec:meth-tc}. First, we focus on static structures to get insight in particular CT features for G4 quadruplexes and also to have a reference for the  following calculations which are applied to the whole MD trajectory. Later on, we are interested in ensemble averages, for single snapshots or static structure cannot elucidate the CT process in DNA. Subsequently, we will compare the CT properties of G4-DNA  with those of double-stranded DNA and distinguish the differences in CT efficiency by means of conformational analysis. Additionally, we determine the effects of the MM environment and in particular of the central ions on CT in G4 quadruplexes.

\subsubsection{Static Structures}
\begin{figure}
\includegraphics{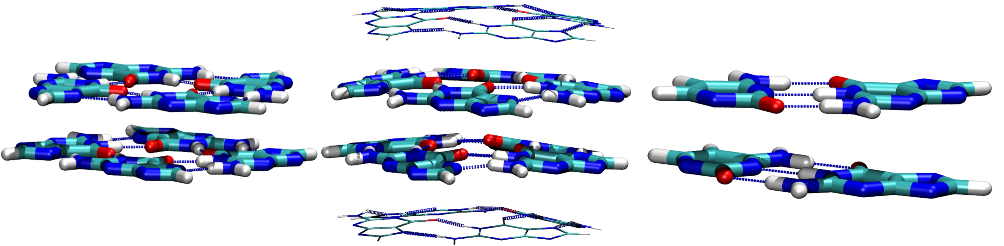}
\caption{Static structures, left: Idealized G4-dimer (ideal), 30\textdegree{} twist and 3.4 \AA{} rise; middle: G4 tetrads of crystal structure 244D (244d); right: double-stranded idealized B-DNA (G$_2$), i.e. two stacked G-C base pairs (36\textdegree{} twist and 3.4 \AA{} rise).}\label{fig:static-strucs}
\end{figure}

\begin{figure}\centering
\includegraphics{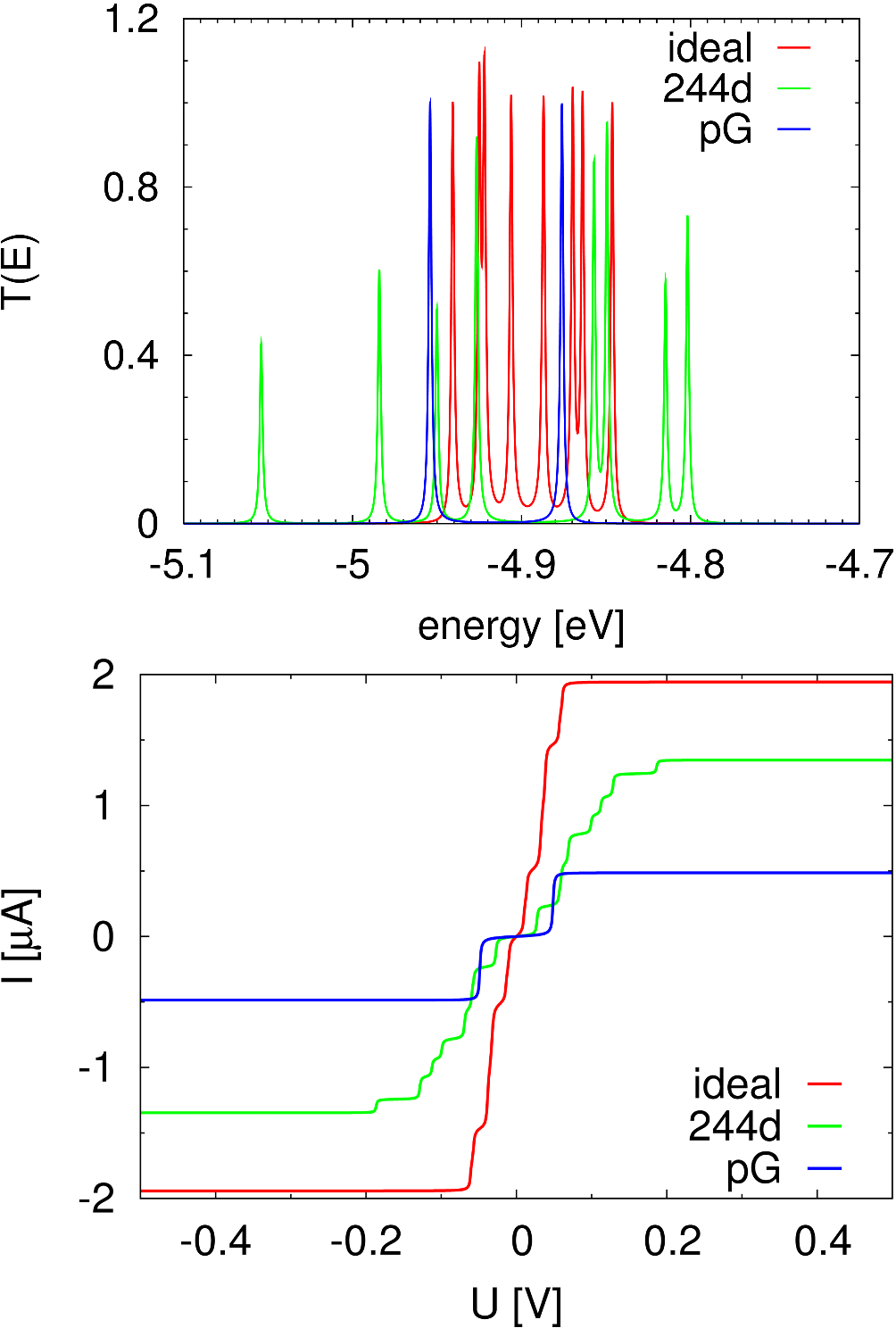}
\caption{Charge transport in static structures from Fig.~\ref{fig:static-strucs}, comparison of transmission function $T(E)$ and current $I(U)$ for idealized G4-dimer, the two central tetrads of crystal structure 244D and poly(G) (G$_2$). Note, only the stacked G bases are considered, whereas the cytosine bases and the backbone composed of sugar and phosphate groups are neglected.}\label{fig:g4-trans-ideal}
\end{figure}

Primarily, we compare T(E) and I(U) for three static structures: i) idealized G4-Dimer as used in Tab.~\ref{tab:hop-g4}, ii) the two central tetrads of the crystal structure 244D and iii) double-stranded idealized B-DNA (G$_2$), i.e. two stacked G-C base pairs (36\textdegree{} twist and 3.4 \AA{} rise). As can be seen from Fig.~\ref{fig:static-strucs} there are only slight structural differences between the idealized G4-dimer and the stacked tetrads of the crystal structure 244D. Fig.~\ref{fig:g4-trans-ideal} exhibits the corresponding transmission functions showing eight resonances due to the eight G sites of the transport model. However, since the idealized G4-dimer has a very regular structure these eight peaks lie in a very narrow energy range of about 0.15 eV and have similar high transmission values of at least 0.9. On the other hand, the eight peaks of the two central tetrads of 244D lie in a broader energy range of about 0.26 eV and have lower transmission values ranging from 0.4 to 0.8. This indicates, how small structural deviations lead to a broadening and a decrease of the transmission function. Nevertheless, we know from the values in Tab.~\ref{tab:hop-g4} that the T1 and T2 couplings are much larger for the crystal structure than for the idealized structure, thus the cause for decreased conductivity has to be the onsite energy disorder.

Fig.~\ref{fig:g4-trans-ideal} also shows the corresponding current-voltage characteristics. The current for the G4 structures is significantly larger than for G$_2$. This is what we would expect, because we compare two idealized molecular units, one with two relevant transmissive states and the other with eight within the relevant energy window. Note that in all current calculations shown in this work, the Fermi energy is artificially placed as the mean value of onsite energies for each snapshot respectively.

Generally, the conductance for the idealized double-stranded poly(G) DNA would become twice as large once the cytosine states are included, though it is well established that the C states have only a minor impact in hole transfer due to their higher oxidation potential than G.

\subsubsection{Influence of Dynamics and Environment}
Previous studies revealed that idealized static structures cannot exhibit reasonable CT properties in DNA, since dynamical as well as environmental effects were shown to be crucial ~\cite{Voityuk2004ACIE,Berlin2008JPCC,Voityuk2006CPL,Voityuk2007CPL,Grozema2008}. Dramatic effects are induced by dynamic disorder, for it suppresses CT in homogeneous sequences on the one hand, but can also enhance CT in heterogeneous (random) sequences on the other. Moreover, only a minority of conformations appears to be CT-active as has been indicated in Ref.~\cite{woiczi2009}. Therefore, neglecting these significant factors or assuming purely random distributions for dynamical disorder leads to a considerable loss of a vital part of the CT in DNA. On this account the CT properties are evaluated for every snapshot along classical MD trajectories which then leads to ensemble averaged quantities, that is the average  transmission function $\left<T(E)\right>$ and the current $\left<I(U)\right>$.

\begin{figure*}\centering
\includegraphics{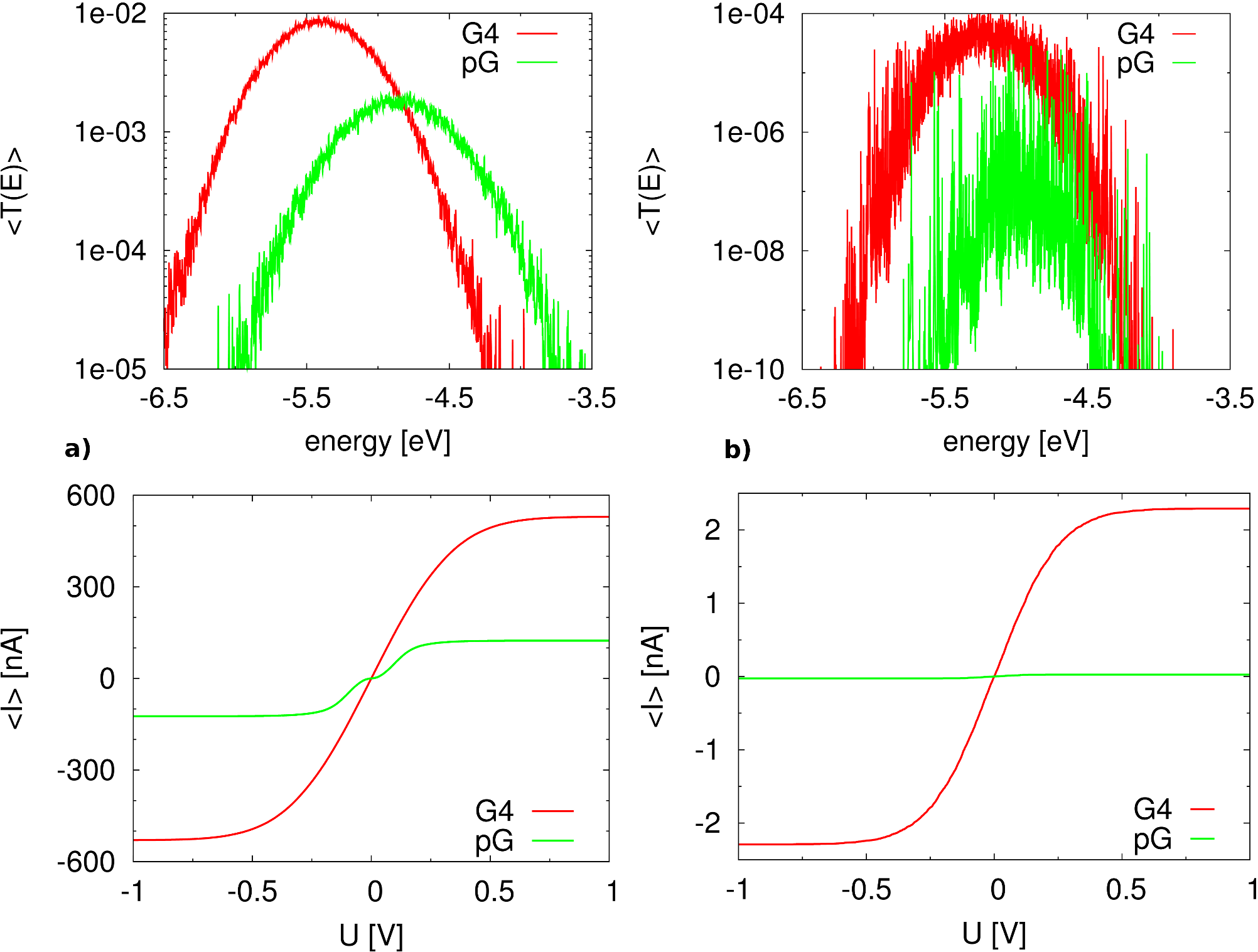}
\caption{Average transmission $\left<T(E)\right>$ and current $\left<I\right>$ obtained from MD simulation: Comparison between G-quadruplex and double-stranded poly(G) DNA. a) (G$_{2}$)$_4$ (Na$^+$) and G$_{2}$, i.e. the two central tetrads (base pairs) of (TG$_4$T)$_4$ (5'-TGGGGT-3'), respectively. b) (G$_{8}$)$_4$ (Na$^+$) and G$_{8}$. Generally, the two last tetrads (base pairs) at the 5' and '3 end are not considered for CT calculations to avoid end effects, although the simulations were done with 12 tetrads and base pairs for G4 and dsDNA, respectively.}\label{fig:compare-dsdna1}
\end{figure*}

In Fig.~\ref{fig:compare-dsdna1} $\left<T(E)\right>$ and $\left<I(U)\right>$ are shown for both quadruplex molecules
 (with central sodium ions), the two central tetrads of (TG$_4$T)$_4$ (a) and (G$_8$)$_{4}$ (b) as well as their corresponding
 double-stranded poly(G) analogues G$_2$ and G$_8$\cite{note1}. Because of the substantial onsite energy fluctuations of about 0.4 eV due to the DNA  environment and dynamics, the transmission function reveals large broadening for both DNA species. The transmission 
maxima for the quadruplexes are shifted to lower energies by about 0.3 eV due to the presence of the central sodium ions.
 Basically, the average transmission is strongly reduced compared to the idealized static structures in Fig.~\ref{fig:g4-trans-ideal}. Nevertheless, the maximum of $\left<T(E)\right>$ for the two central tetrads in (TG$_4$T)$_4$ is almost five times larger compared to G$_2$. Accordingly, the maximum current is about 4.4 times larger 
in the quadruplex. 

Considering the octamers in Fig.~\ref{fig:compare-dsdna1}(b) the conductance difference between G4 and poly(G) even increases. Here, the average transmission in the relevant energy range for the quadruplex is to a great extent two orders of magnitude larger than those for the poly(G) sequence. The poly(G) spectra shows much larger spikes at certain energies. These spikes indicate the strong impact of dynamics and may be explained by the existence of few charge transfer active conformations, which dominate the average transmission, i.e. few conformations, which feature a high transmission. This may indicate that even longer sampling (more than 30 ns) would be required to converge the spectra. Notwithstanding, it also reveals that in dsDNA the average current is dominated to a much larger degree on few non-equilibrium structures, as concluded from our earlier work~\cite{woiczi2009}. As a result, the average current $\left<I(U)\right>$ for (G$_8$)$_{4}$ is almost two orders of magnitude larger than for G$_8$ suggesting that the enhancement of CT in G4 with respect to poly(G) might grow with increasing DNA length. 

Thus, additional sets of CT calculations for longer DNA species with 14 and 20 tetrads (base pairs), respectively, are carried out. The corresponding data was obtained from MD simulations of a quadruplex (G$_{30}$)$_4$ (including central Na$^+$), and double-stranded DNAs G$_{30}$ and a heterogeneous sequence containing 31 sites~\cite{Scheer-Sequence} recently used by Scheer and co-workers in a CT-measurement~\cite{Kang2010} (Note, for CT calculations only the 14 (20) central sites are used). As expected, the transmission strongly decreases for both DNA species. However, in G4 this effect is not as strong as in dsDNA. For instance the current for (G$_{14}$)$_4$ is only about 1 order of magnitude lower than for (G$_{8}$)$_4$, although expanding to (G$_{20}$)$_4$ the current drops significantly additional 3 orders of magnitude. In contrast, $\left<I(U)\right>$ for poly(G) and the ``Scheer'' sequence decreases by more than 10 orders of magnitude by increasing the number of base pairs from 14 to 20. This indicate there is a much stronger distance dependence of CT in dsDNA, hence the notion of coherent CT for longer molecular wires might be considerably more likely in G4 than in dsDNA. Admittedly, the Landauer formalism used in this work performs well for short DNA species (less than 10 sites), where the transport is assumed to be at least partially coherent. On the other hand, it clearly fails for long DNA sequences, hence the CT results for the longer molecules should be interpreted only qualitatively and with caution. For instance, the currents obtained for the 14mer and 20mer of both dsDNA molecules poly(G) and the Scheer sequence are orders of magnitude smaller than pA which is far beyond any measurable range. The complete data is given in Fig.~S3 in Ref.~\cite{supp}.
 
\subsubsection{Analysis of CT Differences in G4 and dsDNA}\label{sec:ana-g4-ct}

The significant conductance difference of G4 and dsDNA may not be attributed to the fact that G4 is composed of four poly(G) like wires. To analyze this further the CT in (G$_{8}$)$_4$ with central sodium ions (full MD) is compared to various models, in which i) only intra- and interstrand\cite{note2} couplings are non-zero (intra+inter) and ii) the quadruplex is separated into its 4 single strands, for which the intrastrand transport is calculated independently and added up afterwards ($\Sigma$4s). Furthermore the results are analyzed with reference to G$_8$ (4xpG). Note, for comparison the CT quantities of poly(G) are multiplied by 4.
 
\begin{figure*}\centering
 \includegraphics{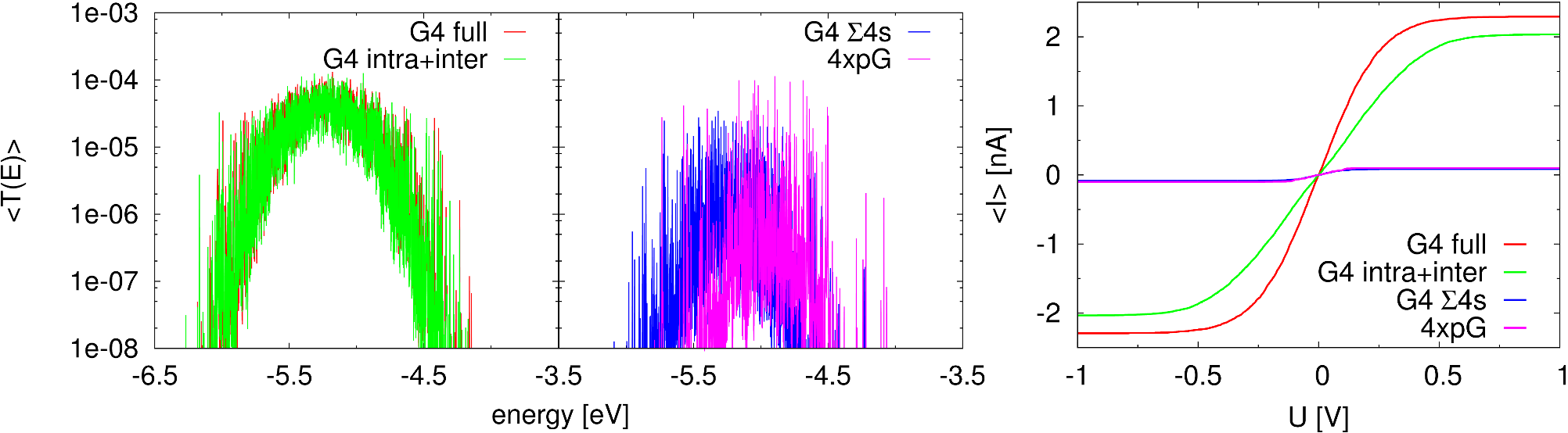}
\caption{Effect of electronic couplings on CT in (G$_{8}$)$_4$ (Na+): Average transmission and current calculated i) for the full time series of couplings obtained from MD (full), ii)  only intra- and interstrand couplings are non-zero (intra+inter), iii) the quadruplex is separated into its 4 single strands ($\Sigma$4s), for which the intrastrand transport is calculated independently and added up afterwards, and iv) for G$_{8}$ multiplied by 4 (4xpG).}\label{fig:transport-fullvsintrainter}
\end{figure*}

As appears from Fig.~\ref{fig:transport-fullvsintrainter}, the average transmissions for the two models: full MD and intra+inter in the spectral support region clearly reveal the largest plateaus, also showing similar peak structures. However, both quadruplex models exhibit similar moderate fluctuations. By contrast, the sum of the four single G4 strands ($\Sigma$4s) shows much larger fluctuations, comparable with those for G$_8$ (4xpG). Moreover, the average transmission for $\Sigma$4s is significantly reduced and is even slightly lower than 4xpG. Interestingly, in $\Sigma$4s are barely CT-active conformations, i.e. single dominating peaks like in 4xpG which sometimes even outreach the maximum transmission of the quadruplex (full MD). This might reflect the structural differences between quadruplexes and dsDNA, since the four strands in the quadruplex are not as flexible as those in poly(G) (see also RMSD fluctuations in Fig.~\ref{fig:rmsd2}), hence in poly(G) the structural phase space is much larger and therefore, several high-transmissive structures can arise. On the other hand, the more rigid quadruplex covers only a smaller conformational phase space which indeed ensures a large number of structures showing moderate CT properties for each single strand, respectively. This clearly indicates, that the most important factor for the enhanced conductance in G4 are the interstrand couplings between the 4 strands in the quadruplex. Thus, if there are conformations in which the 4 isolated channels are not transmissive, there is a considerable probability that CT might occur via coupling between the individual stands. As can be extracted from Tab.~\ref{tab:hop-g4} those interstrand couplings are sufficiently large with about 0.01-0.02 eV. As a consequence, at each snapshot there is a substantial amount of pathways over which the CT might occur through the quadruplex. These findings are supported by the I-V characteristics in Fig.~\ref{fig:transport-fullvsintrainter}. The current for the intra+inter model almost matches that of the full MD with 2.0 and 2.3 nA, respectively. If we switch off the interstrand couplings in the quadruplex ($\Sigma$4s) the current will drop down to 0.08 nA which is more than order of magnitude smaller. A slightly larger current of 0.11 nA is obtained for 4xpG.

\begin{figure}\centering
 \includegraphics{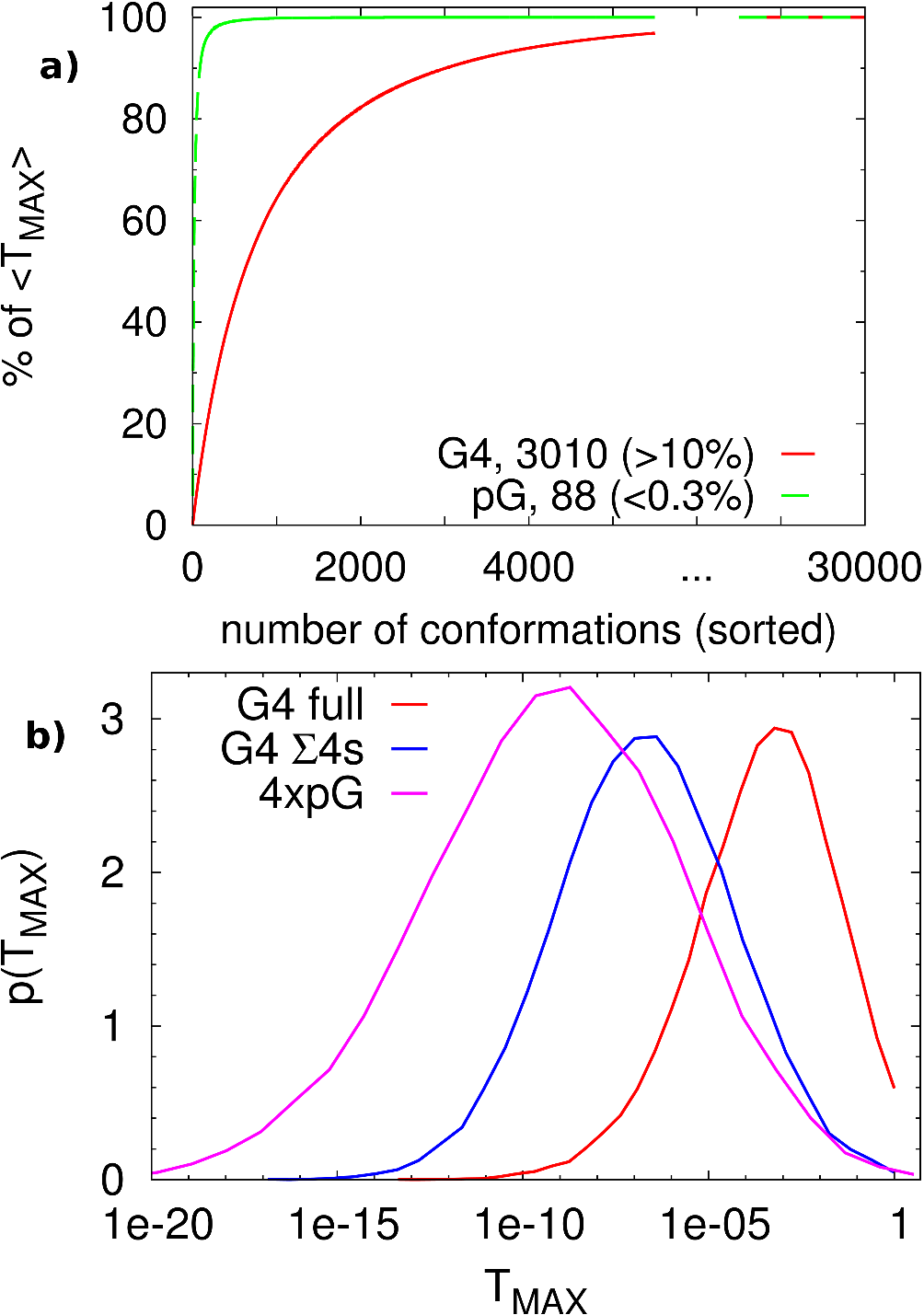} 
\caption{Conformation analysis for (G$_{8}$)$_4$ (Na$^+$) and G$_{8}$, a): number of conformations that make up 90\% of the average transmission maximum $\left<T_{MAX}\right>$, b): probability distribution functions (PDFs) of $T_{MAX}$ for (G$_{8}$)$_4$ (Na$^+$) and the two models 4xpG and $\Sigma$4s as used in Fig.~\ref{fig:transport-fullvsintrainter}.}
\label{fig:transport-conf-G4vspG}
\end{figure}

For further insights into the different CT properties between G4 and dsDNA we make use of conformation analysis. For that purpose, we investigate the amount of conformations which dominate the average CT. Additionally, the distribution of transmissions for the multitude of conformations is explored for both (G$_{8}$)$_4$ and G$_8$. The results are presented in Fig.~\ref{fig:transport-conf-G4vspG}. Panel a) evidently indicates that there are substantially more conformations contributing to the average CT in G4 than in poly(G). Consequently, virtually every 10th G4 conformation is CT-active (about 3000 out of 30,000), whereas  only 128 (again out of 30,000) single non-equilibrium structures characterize the average CT in poly(G).

Second, as demonstrated in the transmission probability distribution functions (PDFs) in Fig.~\ref{fig:transport-conf-G4vspG}b), the distribution width for the G4 quadruplex is considerably narrower compared to 4xpG and $\Sigma$4s. Note the x-axis is scaled logarithmic. In addition, the G4 PDF is significantly shifted to higher transmission. This underscores that the majority of poly(G) structures is not transmissive, yet a few single conformations are responsible for the average CT. On the other hand, for $\Sigma$4s the major part of conformations reveal higher transmission than in 4xpG, but the single dominating conformations are missing. Clearly, this explains that these single high-transmissive poly(G) conformations are the reason for a better conductivity in 4xpG compared to $\Sigma$4s. The higher average T1 values in dsDNA are due to few highly conducting conformations. The smaller average of the G4 T1 couplings therefore resembles a more stable structure, however, not leading to a higher conductance, as could be argued beforehand. The advantage of G4 over dsDNA is due to the existence of non-negligible interstrand couplings in G4. The amount of high-transmissive structures is remarkably increased compared to double-stranded poly(G) DNA.

\begin{figure}\centering
 \includegraphics{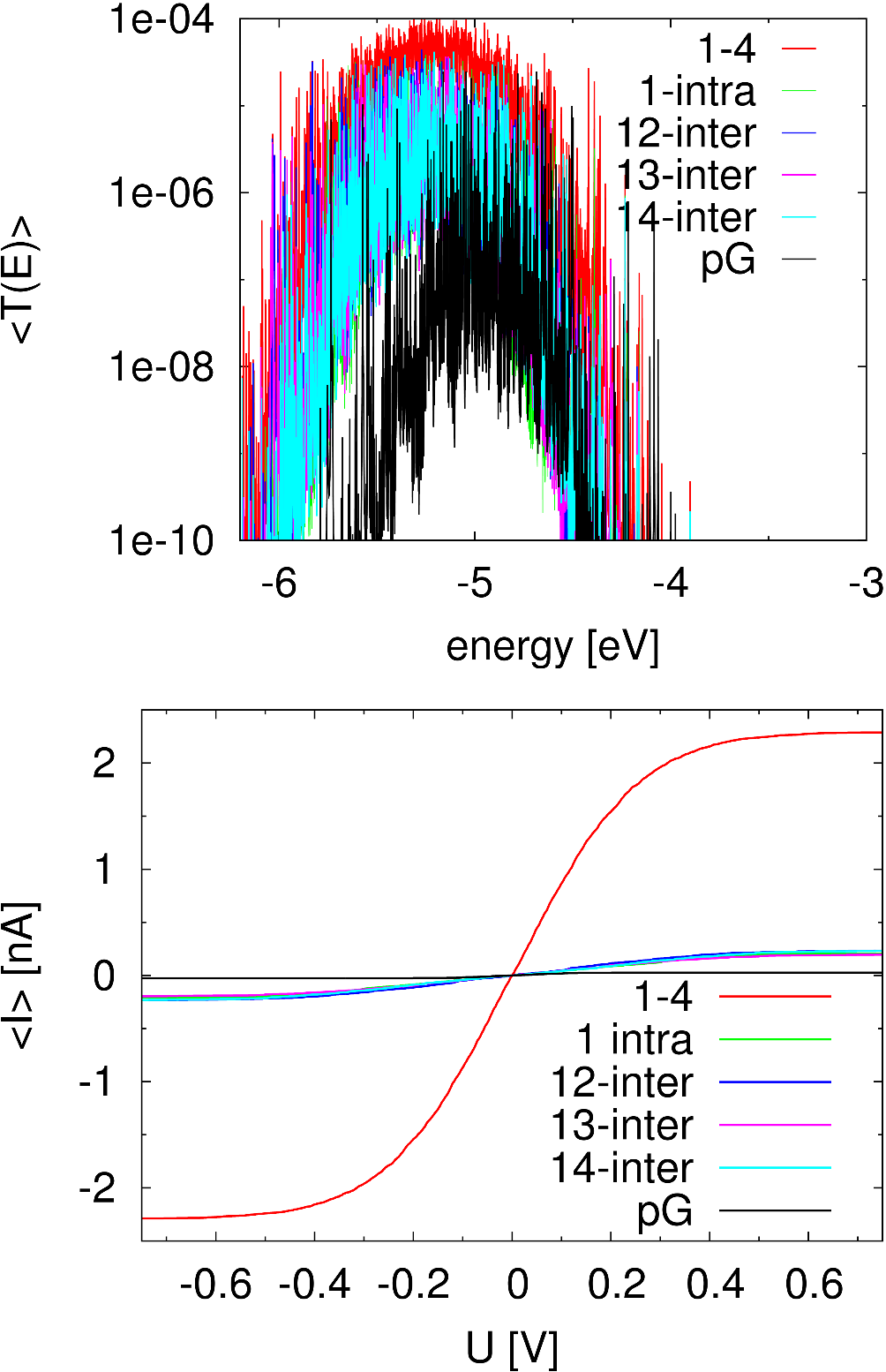}
\caption{Effect of contacts on CT in (G$_{8}$)$_4$ (Na$^+$): Average transmission and current calculated for various contact models i) all 4 strands are connected to the left and right electrodes, respectively (1-4) ii) only one strand is connected to both contacts (1-intra), iii-v)  one strand is calculated to the left electrode while one of the remaining is contacted to the right one (12-inter, etc.), and vi) comparison with poly(G).}\label{fig:transport-contact-ana}
\end{figure}

\paragraph{Contact Effects}
In most of the conductance experiments for DNA the molecules are only connected with one strand to the respective left and right contacts. For instance, in a very recent experiment by Scheer {\textit{et al.}} a single stranded nucleotide with sequence 5'-(T$^*$G$_{3}$[TTAGGG]$_3$T$^*$)-3' (T$^*$ denotes modified thymine residues), which is known to form stable quadruplexes, has been attached between two contacts~\cite{Scheer-g4-2009}. Therefore, the question arises whether the CT in an all-parallel stranded quadruplex differs if only 2 or even 1 strand of the quadruplex are coupled to the left and right contacts, respectively. For that purpose, the CT in (G$_{8}$)$_4$ (Na+) is calculated for various contact models: i) all 4 strands are connected to the left and right electrodes, respectively (1-4), ii) only one strand is connected to both contacts (1-intra) and iii-v)  one strand is attached to the left electrode while one of the remaining is contacted to the right one (12-inter, etc.). As can be seen in Fig.~\ref{fig:transport-contact-ana}, reducing the number of connected strands from 4 to 1 leads to an decrease in $\left<T(E)\right>$ for the relevant energy range of about 1 order of magnitude. However, the transmission for the 1-intra and inter models is very similar indicating a minor significance for which strand or strands are connected to the contacts. This finding is also supported by the I-V characteristics, for the current ranges around 0.2 nA for these models which is roughly one order of magnitude smaller as though all 4 strands are attached (1-4 model).

 Interestingly, there seems to be an increase in the transmission fluctuations if the quadruplex is contacted through only one strand at each end. Notwithstanding, the transmission plateau for the 1-stranded (intra and inter) contacted models is still about 1.5 orders of magnitude larger compared to those for poly(G) resulting in an average current which is again 1 order of magnitude smaller with 0.025 nA. Thus our results suggest that independent on the various contact linking schemes of G4 and dsDNA one might expect a higher conductivity for the quadruplex. Certainly, the optimal conductance for all-parallel stranded G4-DNA is ensured if all four strands are coupled to the contacts.
  
\begin{figure*}\centering
 \includegraphics{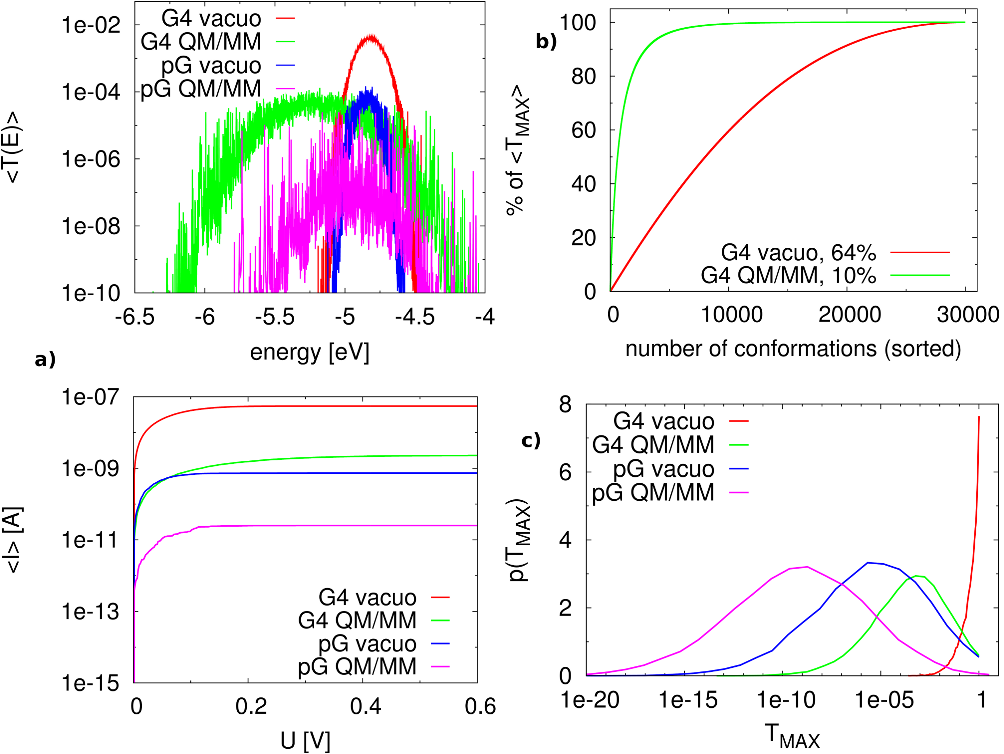}
\caption{Influence of MM environment on CT: Comparison between (G$_{8}$)$_4$ (Na$^+$) and G$_{8}$. a) average transmission and current, b) number of conformations that make up 90\% of the average transmission maximum $\left<T_{MAX}\right>$ and c) probability distribution functions (PDFs) of $T_{MAX}$.}\label{fig:transport-envvsvacuo}
\end{figure*}

\paragraph{Effect of DNA Environment on Transport}
As discussed above, the major part of the dynamical disorder is induced by the QM/MM environment, i.e. the last term in Eq.~\ref{eq:QMMM} which is built of the MM charges of DNA backbone, solvent and counterions. Previous results have indicated that the disorder due to the DNA environment might not only suppress CT in homogeneous sequences like poly(G), rather it is able to enhance CT in random sequences like the heterogeneous Dickerson dodecamer~\cite{woiczi2009}. Very recent experiments by Scheer {\textit{et al.}} confirmed this notion, since they found the current for a 31 base pair sequence~\cite{Scheer-Sequence} to be 2 orders of magnitude smaller in vacuo than in aqueous solution~\cite{Kang2010}. Therefore, it might be interesting to investigate, what happens if we switch of the environment for CT in G4-DNA, also with respect to the effect in dsDNA. As can be seen in Fig.~\ref{fig:transport-envvsvacuo}(a), the transmission maximum for (G$_{8}$)$_4$ in vacuo is about 2 orders of magnitude larger than with the QM/MM environment. Besides, its broadening and also the fluctuations are significantly reduced and the plateau is shifted to higher energies due to the neglect of the electrostatic interaction with the sodium ions within the quadruplex. The transmission for poly(G) in vacuo shows basically the same features, for the broadening is likewise strongly reduced and the maximum is located in the same energy range as for G4 in vacuo, although it is 2 order of magnitude smaller. In general, the transmission for G4 including the QM/MM environment clearly reveals the largest broadening which might indicate that the central sodium ions have an additional strong influence on the dynamic disorder due to longitudinal mobility within the quadruplex, which is not the case in dsDNA. Despite all that, the current at high voltages is larger for G4 with QM/MM than for poly(G) in vacuo.  

Basically, for heterogeneous sequences (not shown here) a reduced transmission in vacuo is found, which is caused by energy gaps between A and G states. Notwithstanding, for DNA molecules with no static energy gaps like double-stranded poly(G) and G4-DNA (both with uniform DNA bases), the QM/MM environment is most likely to increase the dynamical disorder, thus will suppress CT compared to the vacuo model. As a result, there is no significant difference in the effect of the DNA environment on CT for G4 and poly(G) DNA. This is also underscored by conformation analysis given in Panel b) and c) in Fig.~\ref{fig:transport-envvsvacuo}, which indicate that there are considerably more CT-active conformations in vacuo than with QM/MM environment for both DNA species G4 and poly(G). Moreover, for G4 in vacuo nearly every conformation appears to be high-transmissive, since the average maximum transmission $\left<T_{MAX}\right>$ increases almost linearly with the number of conformations. Thus, the CT in vacuo is only marginally affected by single non-equilibrium conformations, rather the whole ensemble of G4 conformations seems to be CT active which is demonstrated in the PDF of transmission. 

\subsubsection{Effect of Channel ions on Transport}
\begin{figure}\centering
 \includegraphics{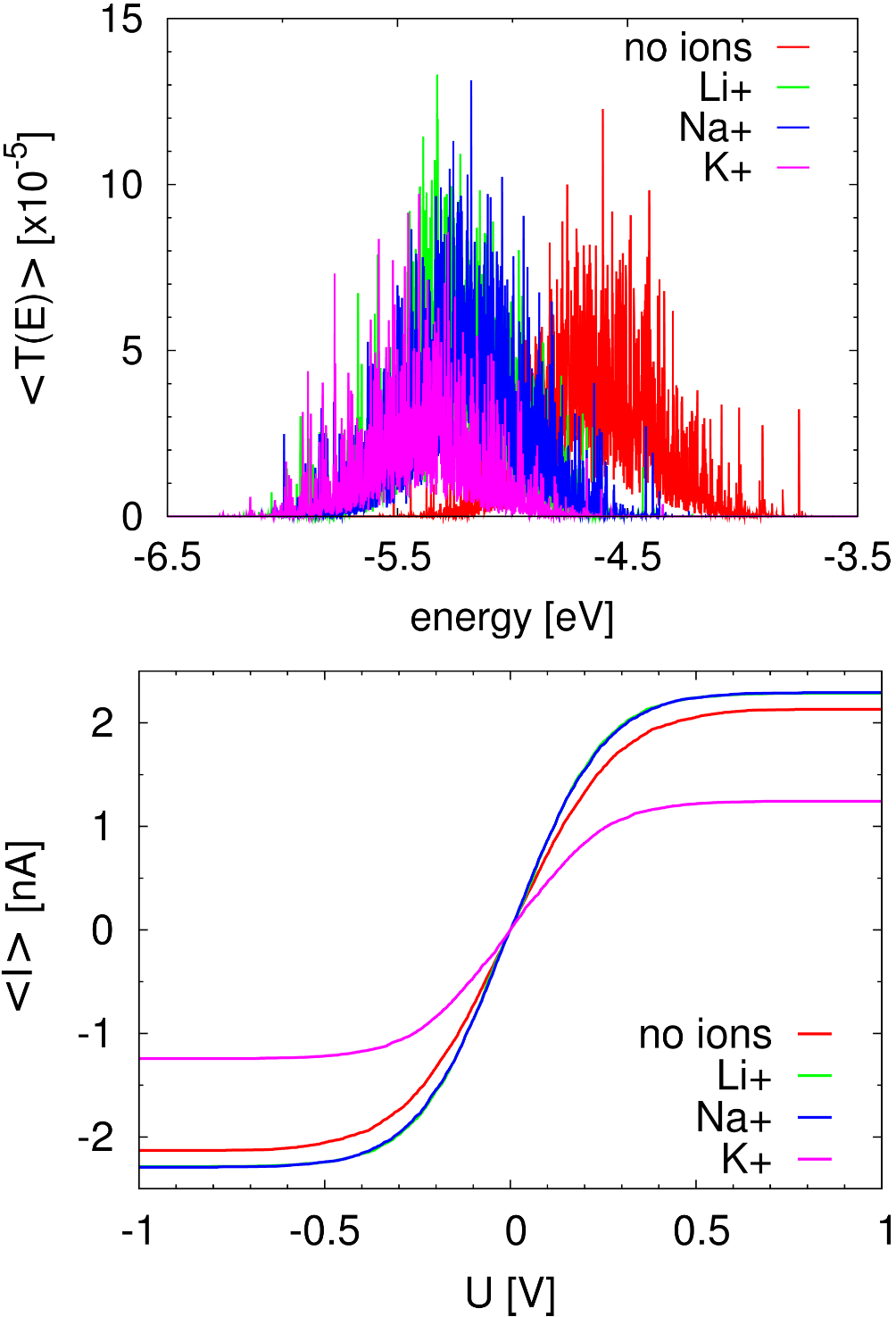}
\caption{Influence of central ions on CT: Calculation of the average transmission and current for (G$_{8}$)$_4$ in absence and presence of monovalent central ions Li$^+$, Na$^+$ and K$^+$.}\label{fig:transport-ion-ch}
\end{figure}

One last issue remains when considering CT in G4-DNA that is the effect of ions within the quadruplex channel. The structural influence of these ions has already been addressed in detail in Sec.~\ref{subsec:res-md}. As could be seen in Sec.~SII.B in Ref.~\cite{supp} we did not find the central ions to contribute states in the relevant energy range for hole transfer in G4, therefore, the effect will only be investigated electrostatically. In Fig.~\ref{fig:transport-ion-ch} the average transmission and current is shown for the quadruplex simulations of (G$_{8}$)$_4$ in absence and presence of either lithium, sodium and potassium ions. Obviously, the transmission maximum is only marginally affected by the presence of different types of central ions. Furthermore, as expected the transmission function is shifted to lower energies if central ions are present. However, for potassium the transmission is to be found slightly reduced, also the broadening is not as large as for the other species. This is also reflected in the PDF of transmission maxima which can be found in Fig.~S4 in Ref.~\cite{supp}. 

As a consequence, the average current for the simulation with central potassium ions is half as large as for those with lithium and sodium ions which might be attributed to the different mobilities of Li$^+$ and Na$^+$ compared to K$^+$. Interestingly, there is no significant difference for CT in absence and presence of ions, although it is known from Sec.~\ref{subsec:res-md} that the G4 molecules without central ions exhibit significant destabilization. Once more this supports the notion that the enhanced conductance in G4 may not exclusively be explained in terms of higher structural rigidity, i.e. less dynamical disorder, rather it is the multitude of CT pathways via interstrand couplings that brings on an increased number of high-transmissive conformations. Apparently, those interstrand and in-plane couplings T2 and T3 are not altered by the presence of central ions, i.e. by a more rigid quadruplex structure.

\section{Discussion and Conclusion}
In this work, we have investigated the conductivity of G4 with respect to dsDNA using classical MD simulations, combined QM/MM methods to compute CT parameters and Landauer theory to compute transmission and I-U characteristics. These properties have been evaluated for phase-space ensembles of G4 and dsDNA structures.

The approach adopted here to calculate the transport characteristics at a given time (Landauer theory) has clear limitations related to the underlying assumption of tunneling transport. Though this may be an efficient pathway over very short segments of the molecules under study, its validity becomes questionable with increasing length. This becomes quite evident when investigating sequences as used in recent experiments e.g. by Scheer and co-workers~\cite{Kang2010}, where our methodology would predict currents order of magnitude smaller than those found in the experiments.  We are currently developing a methodology in order to describe transport in complex materials in a more general way~\cite{Gutierrez2009,Gutierrez2010}.

Nevertheless, the Landauer calculations reported here give interesting insight into particular properties of G4, especially when compared to dsDNA. First of all, G4 with central ions is structurally much more stable than dsDNA and has more $\pi$-contacts along the chain \cite{CCohen2007}. At first sight, one may argue that this leads to improved electrical conduction, however, it turns out that the large fluctuations found for dsDNA lead to highly conducting structures, which dominate the transport. Therefore, a dsDNA molecule conducts better than one strand of G4, or equivalently, the higher conduction of G4 is not due an increased structural stability of its single strands. The phase space of the single strands still contains a vast majority of conformations, which are not CT active. It is the ability of G4 to allow for a large number of conformations due to the interstrand couplings T2 and T3 which makes this species better conducting. If the pathway along one strand is blocked, e.g. when one T$_{ij}$ along the chain vanishes, many other conduction channels may be viable due to interstrand hopping. At the end, G4 has a much larger number of CT active conformations than four dsDNA (poly(G)) offers. This is the basis of the higher conductivity of G4. This advantage is even maintained, when contacted only at one G-site, instead of four sites.

As a second point, dsDNA or G4 may be exerted to strain due to the contacting procedure. Here, clearly the higher stability of G4 may help to maintain a conducting conformation, while the conduction in dsDNA may be much more easily disrupted~\cite{Scheer-g4-2009}.

In a recent work, we have predicted dsDNA with non-uniform sequences (i.e. not poly(G) or poly(A)) to conduct better in solution than in gas phase. This is due to the larger structural fluctuations in solvent, which introduce conducting structures~\cite{woiczi2009}. This finding has been confirmed by a recent experiment~\cite{Kang2010}. For homogeneous sequences like poly(G) or G4 however, a higher conductance in gas phase should be found~\cite{note3}. Finally, our results suggest that the presence of central metal ions within the quadruplex has only a marginal impact on hole transfer in G4, although they are vital for the stability and rigidity of G4-DNA.

\section{Acknowledgment}
This work has been supported by the Deutsche Forschungsgemeinschaft (DFG) within the Priority Program 1243 ``Quantum transport at the molecular scale'' under contract EL 206/5-2 and CU 44/5-2, by the Volkswagen Foundation grant Nr.~I/78-340 and by the European Union under contract IST-029192-2.



\begin{thebibliography}{10}

\bibitem{Boon2003}
E.~Boon, A.~Livingston, N.~Chmiel, S.~David, and J.~Barton,
\newblock {\em Proc.\ Natl.\ Acad.\ Sci.\ USA} {\bf 100}, 12543 (2003).

\bibitem{keren03}
K.~Keren, R.~S. Berman, E.~Buchstab, U.~Sivan, and E.~Braun,
\newblock {\em Science} {\bf 302}, 1380 (2003).

\bibitem{braun98}
E.~Braun, Y.~Eichen, U.~Sivan, and G.~Ben-Yoseph,
\newblock {\em Nature} {\bf 391}, 775 (1998).

\bibitem{pompe99}
M.~Mertig, R.~Kirsch, W.~Pompe, and H.~Engelhardt,
\newblock {\em Eur. Phys. J. D} {\bf 9}, 45 (1999).

\bibitem{Rothemund2006}
P.~Rothemund,
\newblock {\em {Nature}} {\bf {440}}, 297 ({2006}).

\bibitem{Shih2004}
W.~Shih, J.~Quispe, and G.~Joyce,
\newblock {\em {Nature}} {\bf {427}}, 618 ({2004}).

\bibitem{Shih2009}
S.~M. Douglas, H.~Dietz, T.~Liedl, B.~Hoegberg, F.~Graf, and W.~M. Shih,
\newblock {\em {Nature}} {\bf {459}}, 414 ({2009}).

\bibitem{storm01}
A.~J. Storm, J.~V. Noort, S.~D. Vries, and C.~Dekker,
\newblock {\em Appl.\ Phys.\ Lett.} {\bf 79}, 3881 (2001).

\bibitem{porath00}
D.~Porath, A.~Bezryadin, S.~D. Vries, and C.~Dekker,
\newblock {\em Nature} {\bf 403}, 635 (2000).

\bibitem{yoo01}
K.-H. Yoo, D.~H. Ha, J.-O. Lee, J.~W. Park, J.~Kim, J.~J. Kim, H.-Y. Lee,
  T.~Kawai, and H.~Y. Choi,
\newblock {\em Phys.\ Rev.\ Lett.} {\bf 87} (2001).

\bibitem{tao04a}
B.~Xu, P.~Zhang, X.~Li, and N.~Tao,
\newblock {\em Nano Lett.} {\bf 4}, 1105 (2004).

\bibitem{cohen05}
H.~Cohen, C.~Nogues, R.~Naaman, and D.~Porath,
\newblock {\em Proc.\ Natl.\ Acad.\ Sci.\ USA} {\bf 102}, 11589 (2005).

\bibitem{Kang2010}
N.~Kang, A.~Erbe, and E.~Scheer,
\newblock {\em {App. Phys. Lett. }} {\bf {96}} ({2010}).

\bibitem{tao04b}
X.~Xiao, B.~Q. Xu, and N.~J. Tao,
\newblock {\em Nano Lett.} {\bf 4}, 267 (2004).

\bibitem{Berlin2004}
Y.~A. Berlin, I.~V. Kurnikov, D.~Beratan, M.~A. Ratner, and A.~L. Burin,
\newblock {\em Top. Curr. Chem.} {\bf 237}, 1 (2004).

\bibitem{Grozema2002}
F.~C. Grozema, L.~D.~A. Siebbeles, Y.~A. Berlin, and M.~A. Ratner,
\newblock {\em ChemPhysChem} {\bf 3}, 536 (2002).

\bibitem{Senthilkumar2005}
K.~Senthilkumar, F.~C. Grozema, C.~F. Guerra, F.~M. Bickelhaupt, F.~D. Lewis,
  Y.~A. Berlin, M.~A. Ratner, and L.~D.~A. Siebbeles,
\newblock {\em J.\ Am.\ Chem.\ Soc.} {\bf 127}, 14894
  (2005).

\bibitem{CramerT._jp071618z}
T.~Cramer, S.~Krapf, and T.~Koslowski,
\newblock {\em J. Phys. Chem. C} {\bf 111}, 8105 (2007).

\bibitem{Gutierrez2009}
R.~Guti\'{e}rrez, R.~A. Caetano, B.~P. Woiczikowski, T.~Kuba\v{r}, M.~Elstner,
  and G.~Cuniberti,
\newblock {\em {Phys. Rev. Lett.}} {\bf {102}} ({2009}).

\bibitem{woiczi2009}
P.~B. Woiczikowski, T.~Kuba\v{r}, R.~Guti\'{e}rrez, R.~A. Caetano,
  G.~Cuniberti, and M.~Elstner,
\newblock {\em J.\ Chem.\ Phys.} {\bf 130}, 215104 (2009).

\bibitem{Phan2002}
A.~Phan and J.~Mergny,
\newblock {\em {Nucleic Acids Res.}} {\bf {30}}, 4618 ({2002}).

\bibitem{Williamson1994rev}
J.~Williamson,
\newblock {\em {Ann. Rev. Biophys. Biomol. Struc.}} {\bf {23}}, 703 ({1994}).

\bibitem{Arthanari2001}
H.~Arthanari and P.~Bolton,
\newblock {\em {Chem. \& Biol.}} {\bf {8}}, 221 ({2001}).

\bibitem{Mergny1999}
J.~Mergny, P.~Mailliet, F.~Lavelle, J.~Riou, A.~Laoui, and C.~Helene,
\newblock {\em {Anti-Cancer Drug Design}} {\bf {14}}, 327 ({1999}).

\bibitem{Maizels2006}
N.~Maizels,
\newblock {\em Nature Struc. \& Mol. Biol.} {\bf 13}, 1055 (2006).

\bibitem{Sen1988}
D.~Sen and W.~Gilbert,
\newblock {\em Nature} {\bf 334}, 364 (1988).

\bibitem{Sen1990}
D.~Sen and W.~Gilbert,
\newblock {\em Nature} {\bf 344}, 410 (1990).

\bibitem{Sen1992}
D.~Sen and W.~Gilbert,
\newblock {\em Methods in Enzymology} {\bf 211}, 191 (1992).

\bibitem{Hoogsteen1963}
K.~Hoogsteen,
\newblock {\em Acta Crystallographica} {\bf 16}, 907 (1963).

\bibitem{Davis2004}
J.~Davis,
\newblock {\em Angew. Chem. Int. Ed.} {\bf 43}, 668 (2004).

\bibitem{Simonsson2001}
T.~Simonsson,
\newblock {\em Biol. Chem.} {\bf 382}, 621 (2001).

\bibitem{CCohen2007}
H.~Cohen, T.~Sapir, N.~Borovok, T.~Molotsky, R.~Di~Felice, A.~B. Kotlyar, and
  D.~Porath,
\newblock {\em Nano Lett.} {\bf 7}, 981 (2007).

\bibitem{Kotlyar2005}
A.~B. Kotlyar, N.~Borovok, T.~Molotsky, H.~Cohen, E.~Shapir, and D.~Porath,
\newblock {\em Adv. Mat.} {\bf 17}, 1901 (2005).

\bibitem{Borovok2008}
N.~Borovok, T.~Molotsky, J.~Ghabboun, D.~Porath, and A.~Kotlyar,
\newblock {\em {Anal. Biochem.}} {\bf {374}}, 71 ({2008}).

\bibitem{Spackova1999}
N.~\v{S}pa\v{c}kov\'{a}, I.~Berger, and J.~\v{S}poner,
\newblock {\em J.\ Am.\ Chem.\ Soc.} {\bf 121}, 5519
  (1999).

\bibitem{Spackova2001}
N.~\v{S}pa\v{c}kov\'{a}, I.~Berger, and J.~\v{S}poner,
\newblock {\em J.\ Am.\ Chem.\ Soc.} {\bf 123}, 3295
  (2001).

\bibitem{Cavallari2006}
M.~Cavallari, A.~Calzolari, A.~Garbesi, and R.~Di~Felice,
\newblock {\em J.\ Phys.\ Chem.\ B} {\bf 110}, 26337 (2006).

\bibitem{DiFelice2006book}
R.~D. Felice and A.~Calzolari,
\newblock {\em Electronic structure of DNA derivatives and mimics by Density
  Functional Theory}, chapter~20, pp. 485--508,
\newblock Modern Methods for Theoretical Physical Chemistry of Biopolymers,
  Elsevier, 2006.

\bibitem{Calzolari2004}
A.~Calzolari, R.~DiFelice, E.~Molinari, and A.~Garbesi,
\newblock {\em J.\ Phys.\ Chem.\ B} {\bf 108}, 2509 (2004).

\bibitem{Calzolari20023331}
A.~Calzolari, R.~Di~Felice, E.~Molinari, and A.~Garbesi,
\newblock {\em Appl.\ Phys.\ Lett.} {\bf 80}, 3331 (2002).

\bibitem{Calzolari2004557}
A.~Calzolari, R.~Di~Felice, and E.~Molinari,
\newblock {\em Solid State Commun.} {\bf 131}, 557 (2004).

\bibitem{DiFelice20041256}
R.~Di~Felice, A.~Calzolari, and H.~Zhang,
\newblock {\em Nanotechnology} {\bf 15}, 1256 (2004).

\bibitem{DiFelice200522301}
R.~Di~Felice, A.~Calzolari, A.~Garbesi, S.~Alexandre, Soler, and J.M.,
\newblock {\em J.\ Phys.\ Chem.\ B} {\bf 109}, 22301 (2005).

\bibitem{Guo2009}
A.-M. Guo and S.-J. Xiong,
\newblock {\em {Phys. Rev. B}} {\bf {80}} ({2009}).

\bibitem{Guo2010}
A.-M. Guo, Z.~Yang, H.-J. Zhu, and S.-J. Xiong,
\newblock {\em {J.\ Phys.-Condens.\ Matter}} {\bf {22}} ({2010}).

\bibitem{Skourtis2008}
I.~A. Balabin, D.~N. Beratan, and S.~S. Skourtis,
\newblock {\em Phys. Rev. Lett.} {\bf 101} (2008).

\bibitem{Voityuk2004ACIE}
A.~A. Voityuk, K.~Siriwong, and N.~Rösch,
\newblock {\em Angew. Chem. Int. Ed.} {\bf 43}, 624 (2004).

\bibitem{Berlin2008JPCC}
Y.~A. Berlin, F.~C. Grozema, L.~D.~A. Siebbeles, and M.~A. Ratner,
\newblock {\em J.\ Phys.\ Chem.\ C} {\bf 112}, 10988 (2008).

\bibitem{Voityuk2006CPL}
A.~Sadowska-Aleksiejew, J.~Rak, and A.~A. Voityuk,
\newblock {\em Chem. Phys. Lett.} {\bf 429}, 546 (2006).

\bibitem{Voityuk2007CPL}
A.~A. Voityuk,
\newblock {\em Chem. Phys. Lett.} {\bf 439}, 162 (2007).

\bibitem{Grozema2008}
F.~C. Grozema, S.~Tonzani, Y.~A. Berlin, G.~C. Schatz, L.~D.~A. Siebbeles, and
  M.~A. Ratner,
\newblock {\em J.\ Am.\ Chem.\ Soc.} {\bf 130}, 5157
  (2008).

\bibitem{Gutierrez2010}
R.~Guti\'{e}rrez, R.~Caetano, P.~B. Woiczikowski, T.~Kuba\v{r}, M.~Elstner, and
  G.~Cuniberti,
\newblock {\em {New J. Phys.}} {\bf {12}} ({2010}).

\bibitem{Laughlan1994}
G.~Laughlan, A.~Murchie, D.~Norman, M.~Moore, P.~Moody, D.~Lilley, and
  B.~Luisi,
\newblock {\em Science} {\bf 265}, 520 (1994).

\bibitem{link00}
\emph{http://structure.usc.edu/make-na/server.html}.

\bibitem{TIP3P}
W.~L. Jorgensen, J.~Chandrasekhar, J.~D. Madura, R.~W. Impey, and M.~L. Klein,
\newblock {\em J. Chem. Phys.} {\bf 79}, 926 (1983).

\bibitem{Gromacs}
D.~van~der Spoel, E.~Lindahl, B.~Hess, G.~Groenhof, A.~E. Mark, and H.~J.~C.
  Berendsen,
\newblock {\em J. Comput. Chem.} {\bf 26}, 1701 (2005).

\bibitem{parm99}
J.~Wang, P.~Cieplak, and P.~A. Kollman,
\newblock {\em J. Comput. Chem.} {\bf 21}, 1049 (2000).

\bibitem{parmBSC0}
A.~P\'erez, I.~March\'an, D.~Svozil, J.~\v{S}poner, T.~E. Cheatham~III, C.~A.
  Laughton, and M.~Orozco,
\newblock {\em Biophys. J.} {\bf 92}, 3817 (2007).

\bibitem{Kubar2008b}
T.~Kuba\v{r} and M.~Elstner,
\newblock {\em J.\ Phys.\ Chem.\ B} {\bf 112}, 8788 (2008).

\bibitem{Kubar2008a}
T.~Kuba\v{r}, P.~B. Woiczikowski, G.~Cuniberti, and M.~Elstner,
\newblock {\em J.\ Phys.\ Chem.\ B} {\bf 112}, 7937 (2008).

\bibitem{ElstnerPRB1998}
M.~Elstner, D.~Porezag, G.~Jungnickel, J.~Elsner, M.~Haugk, T.~Frauenheim,
  S.~Suhai, and G.~Seifert,
\newblock {\em Phys. Rev. B} {\bf 58}, 7260 (1998).

\bibitem{LoewdinJCP1950}
P.~O. Loewdin,
\newblock {\em J. Chem. Phys.} {\bf 18}, 365 (1950).

\bibitem{mujica:6849}
V.~Mujica, M.~Kemp, and M.~A. Ratner,
\newblock {\em J. Chem. Phys.} {\bf 101}, 6849 (1994).

\bibitem{mujica:6856}
V.~Mujica, M.~Kemp, and M.~A. Ratner,
\newblock {\em J. Chem. Phys.} {\bf 101}, 6856 (1994).

\bibitem{Spackova2004}
E.~Fadrn\'{a}, N.~\v{S}pa\v{c}kov\'{a}, R.~\v{S}tefl, J.~Ko\v{c}a,
  T.~Cheatham~III, and J.~\v{S}poner,
\newblock {\em Biophys. J.} {\bf 87}, 227 (2004).

\bibitem{Ndlebe2006}
T.~Ndlebe and G.~Schuster,
\newblock {\em Org. Biomol. Chem.} {\bf 4}, 4015 (2006).

\bibitem{Scheer-g4-2009}
S.~P. Liu, S.~H. Weisbrod, Z.~Tang, A.~Marx, E.~Scheer, and A.~Erbe,
\newblock {\em Angew. Chem. Int. Ed.} {\bf in press} (2010).

\bibitem{Scheer-Sequence}
5-thiol-dG-GGC GGC GAC CTT CCC GCA GCT GGT ACG GAC

\bibitem{note1}
Note, generally, the two last tetrads (base pairs) at the 5' and '3 end are not considered for CT calculations to avoid end effects, although the simulations were done with 12 tetrads and base pairs for G4 and dsDNA, respectively.

\bibitem{note2}
Only T2 interstrand couplings between adjacent strands are considered. See also the scheme in Fig.~\ref{fig:hop-scheme}.

\bibitem{note3}
Note, that this only applies when the whole structure is homogeneous. If other bases like A or T enter the sequence, the behavior may be the opposite.

\bibitem{supp}
See Supplementary Material at http://dx.doi.org/10.1063/1.3460132 for further data analysis. 
\end{thebibliography}
\end{document}